\documentclass[11pt,a4paper]{article}

\usepackage{appendix}

\usepackage[utf8]{inputenc}
\usepackage[T1]{fontenc}
\usepackage{lmodern}
\usepackage{microtype}
\usepackage[
    backend=biber,
    style=numeric,
    sorting=none
]{biblatex}
\usepackage{geometry}
\usepackage{authblk}
\usepackage{amsmath}
\usepackage{amssymb}
\usepackage{amsthm}
\usepackage{mathtools}

\usepackage{braket}
\usepackage{bm}

\usepackage{graphicx}
\usepackage{booktabs}

\usepackage{hyperref}
\hypersetup{
    colorlinks=true,
    citecolor=blue,
    linkcolor=blue,
    urlcolor=blue
}

\newtheorem{lemma}{Lemma}
\newtheorem{proposition}{Proposition}
\newtheorem{definition}{Definition}

\newcommand{\F}{\mathbb{F}}
\newcommand{\Z}{\mathbb{Z}}
\newcommand{\h}{\mathbb{H}}
\newcommand{\C}{\mathbb{C}}

\title{Perspectives on the Hidden Subgroup Problem over Unitriangular Groups}

\author[1,2]{Damien Poquillon}
\affil[1]{
    Télécom Paris, Institut Polytechnique de Paris, France
}
\affil[2]{
    University of Ottawa, Ottawa, Canada
}

\begin{document}

\maketitle

\begin{abstract}
    We investigate the quantum Hidden Subgroup Problem (HSP) over the
    upper unitriangular group $U_n(\mathbb{F}_p)$.
    While efficient quantum algorithms were discovered for some families of non abelian groups and nil-2 $p$-groups, no efficient algorithms are known for groups of nilpotency class higher than 2. The only techniques known for such groups scale exponentially in the size of the input.
    
    We first study the group $U_4(\mathbb{F}_p)$, the smallest upper
    unitriangular group exhibiting nilpotency class three. We analyze the application of the Pretty Good Measurement approach used for Heisenberg groups and show its limits. We then focus on the more general case $U_n(\mathbb{F}_p)$ to adapt existing techniques to this particular structure. In particular, we derive an algorithm for cyclic subgroups as a subroutine for a general algorithm, although it is not efficient in the size of the group.  
    
    Our results highlight the difficulties introduced by the increasing
    nilpotency class and suggests some directions for an efficient
    quantum algorithm for the HSP over $U_n(\mathbb{F}_p)$.

\end{abstract}


\vspace{1em}

\tableofcontents

\section{Introduction} \label{section: intro}

The Hidden Subgroup Problem (HSP) is a central problem in quantum computing around which important algorithms were made. In particular, Shor's factoring and discrete logarithm algorithms \cite{Shor95} can be formulated in terms of the abelian HSP, which can be efficiently solved using the Quantum Fourier Transform (QFT). This has motivated the study of the HSP over non-abelian groups to understand the limits of quantum algorithms and to investigate potential applications to problems such as lattice problems and graph isomorphism. This extension was successful for specific families of groups like Heisenberg groups (see \cite{KR08}) and certain classes of nilpotent groups (see \cite{ISS07-2}). A quantum algorithm for general nilpotent groups was also derived more recently in \cite{Iv23}, but is only efficient for small characteristics and small nilpotency classes. 
\\
A next possible step is to consider an explicit representation of a nilpotent group to explore what happens for nilpotency class 3 and above. In this paper, we focus on the upper unitriangular group $U_n(\F_p)$ (group of square matrices of size $n$ over $\F_p$ that are upper triangular with ones on the diagonal) which is a very natural example nilpotent of class $n-1$.  The main issue with $U_n(\F_p)$ is that its representation theory is "wild", and one cannot reasonably hope to find a general QFT over the group. This requires an adaptation of the standard method, which could be of great insight for other groups where the same approach fails. 
\\
We first study the HSP on $U_4$, a group of nilpotency class 3. It ressembles a lot the Heisenberg group of size $5$, with just one extra component. One could then try to exploit methods that worked for such groups, like the Pretty Good Measurement approach (originally used in the HSP setting in \cite{BCD05}. We explore this approach in Section \ref{section: U4 general reduction} and show a partial result for hidden subgroups that are "big" enough (i.e. of order larger or equal than $p^3$). We also explain in details in Section \ref{section: U4 for p^2 and cyclic} how this method already has its limits for order $p^2$ hidden subgroups, and why it is inefficient for cylcic and trivial subgroups. This suggests a different approach: exploiting the abelian behavior of $U_4$ on its first superdiagonal. This method comes at the cost of an exponential time in the characteristic of the field, but is promising for bigger groups.
\\
Finally, we consider the HSP over the general group \(U_n(\mathbb F_p)\). We develop an approach based on reducing the problem to instances involving cyclic hidden subgroups. Although this leads to a simpler treatment adapted to the structure of unitriangular groups, we do not obtain an improvement over existing algorithms for nilpotent groups in terms of computational complexity. Nevertheless, this approach provides a framework for studying the problem across the whole family \(U_n(\mathbb F_p)\), and suggests several directions for further investigation.

\section{Basic group definitions and properties}
\label{section:definitions}

Throughout this paper, $p$ denotes a prime number and $\mathbb{F}_p$ the finite field with $p$ elements. 

\begin{definition}[Nilpotent group]
\label{def:nilpotent-group}
A group $G$ is nilpotent if its lower central series
terminates at the trivial subgroup, or equivalently, if it
admits a central series.
The nilpotency class of a nontrivial nilpotent group $G$
is the smallest integer $r$ such that $G_{r+1}:=[G,[G,...[G,G]...] = \{1\}$.
\end{definition}

\begin{definition}[Upper unitriangular groups]
    $U_n(\F_p)$ denotes the upper triangular groups of matrices of size $n \times n$ over $\F_p$ with 1's on the diagonal and 0's bellow it. It is a nilpotent group of class $n-1$.
\end{definition}

\begin{definition}[Heisenberg group] \label{def:Heisenberg}
For any $n\geq 1$, the Heisenberg group
$H_{2n+1}(\mathbb{F}_p)$ is the subgroup of
$U_{n+2}$ consisting of matrices whose potentially
nonzero entries outside the diagonal belong to the
first row or the last column.
It is a $p$-group of order $p^{2n+1}$ and nilpotent of class $n+1$
\end{definition}

For instance, the three-dimensional Heisenberg group is
\[
\h_3=
\left\{
\begin{pmatrix}
1 & b & a\\
0 & 1 & c\\
0 & 0 & 1
\end{pmatrix}
\ \middle|\
(a,b,c)\in\mathbb{F}_p^3
\right\}.
\]

\begin{proposition}[Order of a product of two subgroups]
\label{prop:order-product}
Let $H_1,H_2$ be two subgroups of a finite group $G$.
The set $H_1H_2=\{h_1h_2\mid h_1\in H_1,\ h_2\in H_2\}$ 
has cardinality
\[
|H_1H_2|=
\frac{|H_1||H_2|}{|H_1\cap H_2|}
\leq |G|
\]
Moreover, if $H_1$ or $H_2$ is normal in $G$, then $H_1H_2$ is a valid subgroup of $G$
\end{proposition}

\section{Progress on resolutions of non abelian HSP instances} \label{section: progress}
\subsection[Extension to the Heisenberg group h3]{Extension to the Heisenberg group $\h_3$} {\label{subsection: 3.2}}
The first natural step to extend the standard method for abelian HSP is to consider groups that are almost abelian, like the Heisenberg group $\h_3$. The efficiency of the abelian HSP can still be exploited when studying a non-abelian HSP: given a hidden subgroup $H\leq G$ and a known abelian subgroup $K \leq G$, one can prepare a uniform superposition over $K$ instead of $G$. This is effectively like restricting the hiding function $f$ to the subgroup $K$, or using an oracle implementing $f|_K$. 
Within this abelian subgroup, one can run the standard method to efficiently solve the HSP to obtain generators of $H \cap K$. When such a procedure is being done, we say that we are solving the HSP in the abelian group $K$. 
\\
To make use of this, one should start the resolution of a non abelian HSP problem by studying the subgroup structure of the group, and using the abelian HSP as much as possible to try to reduce the general problem to a simpler case. Ettinger and Hoyer \cite{EH98} found such a reduction for the dihedral group, which proved useful for the construction of Kuperberg's sieve \cite{Kup03}. They showed that solving the HSP for cyclic subgroups of the dihedral group is enough to solve the HSP on the whole group. This method can be adapted to any group of the form $G=A \rtimes \Z_p$, where $A$ is abelian as claimed in Lemma 1 in \cite{BCD05} :
\begin{lemma} {\label{lemma: hsp in A x Zp}} 
    Let $G = A \rtimes_{\varphi} \Z_p$, with A an abelian group and $\varphi:\Z_p \to Aut(A)$ a group homomorphism. Then to solve the HSP on $G$, it is enough to solve the HSP on subgroups that are trivial or of the form $\langle(d,1)\rangle, d\in A/A'$ in the new group $A/A' \rtimes_{\varphi'} \Z_p$, where $A'\times\{0\}$ is a normal subgroup of $G$ and $\varphi'$ is defined by \[\forall (a,k) \in A\times \Z_p, \ \varphi'(k)(a\cdot A')=\varphi(k)(a)\] 
\end{lemma}
The reduction's idea is to run the HSP on the abelian group $A \times \{0\}$. The standard method will yield generators of the group $A' := H \cap (A \times \{0\})$. It can be shown that if $A' \times \{0\}$ is not a normal subgroup of $G$, then $H=A'$. If the group is normal, one can consider the new group $G' = G / A'$ and solve the HSP on this smaller group.
$\h_3 \simeq \Z_p^2 \rtimes \Z_p$ falls in this category. This reduction leaves only the case where the hidden subgroup is cyclic: if it isn't, the remaining problem is abelian. The cyclic case can be handled using the Pretty Good Measurement (PGM) approach developed in \cite{BCD05}. The method combines weak Fourier sampling with a joint measurement on several coset states. More specifically, it constructs a unitary that maps the uniform superposition over the solutions of a system of two equations to a computational-basis state, thereby allowing the hidden parameters to be recovered through an inverse QFT followed by measurement in the computational basis. A detailed explanation can be found in Andrew Childs lecture notes \cite{Chi25}.

The PGM approach also extends to certain metacyclic groups of the form $\Z_n\rtimes\Z_p$ and groups of the form $\Z_p^r\rtimes\Z_p$. Its generalization relies on solving a related problem known as the matrix sum problem, which can be efficiently handled for these particular families. In the dihedral case, this problem is closely related to average-case subset sum, which is believed to be computationally hard. However, this approach does not directly extend to general Heisenberg groups, for which different techniques are used.

\subsection{Extension to the general Heisenberg group and extraspecial groups} {\label{subsection: 3.3}}

A related approach for $\h_3$ was used by D. Bacon in \cite{Bac07}, which is based on another representation-theory related operation: the Clebsch-Gordan transform (CGT). This transform exploits symmetric relations on few coset states, unlike the QFT that only acts on one coset state. Krovi and Rotteler \cite{KR08} generalized this idea to the general Heisenberg group $\h_{2n+1}$ of order $p^{2n+1}$. 
\\
The resolution of the HSP for the more general class of extraspecial $p$-groups was then derived in a different way in \cite{ISS07-1}. We remind that for such groups, the first commutator $G'= [G,G]$ is equal to the center of the group ($G'=Z$) and it is a cyclic group. This work is based on the idea that if one can solve efficiently the HSP for $HG'$, then one can solve efficiently the HSP for $H$. Indeed, suppose we have a quantum algorithm solving the HSP for $HG'$. In one query to the oracle on an element of $G'$, we can test if $G' \subset H$, and in this case, a quantum algorithm for the HSP on $H=HG'$ would describe $H$. 
Otherwise, $H \cap G' = \{0\}$, and in this case, $HG'$ is abelian. So we can run the abelian HSP by restricting the hiding function to $HG'$, which would give the subgroup $H$. 
Moreover, given copies of coset states for $HG'$, there exist a quantum algorithm that can describe $HG'$ in logarithmic time. This comes from the fact that $G$ is extraspecial, so that $G/G'$ is abelian. Then, by considering the quotient $HG'/G' \leq G/G'$, one is left with an abelian HSP to solve. So the main difficulty of the problem is to find a procedure that can produce coset states for $HG'$ from coset states of $H$. 
\\
For extraspecial $p$-groups, one can start by creating a coset state $\ket{aH}$ for a uniformly random $a\in G$. Then, one can create a uniform superposition of elements in $G'$ and multiply the two registers together to obtain the state $\frac{1}{\sqrt{|G'|}}\displaystyle\sum_{z\in G'}\ket{aHz}\ket{z}$. The state we are aiming for is $\frac{1}{\sqrt{|G'|}}\displaystyle\sum_{z\in G'}\ket{aHz}$, but there is no way to completely get rid of the entangled last register. We can try to do a QFT and to measure some outcome $u\in G'$, which would give the state $\ket{aHG_u'} = \frac{1}{\sqrt{p}}\displaystyle\sum_{z \in G'}\omega^{zu}\ket{aHz}$. Using 4 copies of states obtained this way, it is actually possible to produce a purification of the mixed objective state $\rho_{HG'} = \frac{1}{|G|}\displaystyle\sum_{a\in G}\ket{aHG'} \bra{aHG'}$. This procedure can be made efficiently in the size of the input (Theorem 3, \cite{ISS07-1}), and involves the resolution of a quadratic system that depends on the random outcomes $u\in \Z_p$ obtained after the measurement of the last register. Using 4 copies insure that the system has non trivial solutions. This looks like the situation of the Heisenberg resolution, where several copies of coset states were needed to ensure the system of equation had a constant number of solutions.

\subsection[Extension to nil 2 p groups]{Extension to nil-2 $p$-groups} \label{subsection nil2}
The resolution for nil-2 $p$-groups in \cite{ISS07-2} follows the same path as extraspecial ones. For such a nil-2 groups of order $p^n$, the first commutator group is abelian and isomorphic to $\Z_p^d$, for some $d\leq n$. Here, the generation of a purification of the mixed state $\rho_{HG'}$ can still be made efficiently in $n\log(p)$ (Theorem 4, \cite{ISS07-2}). This procedure involves a particular transformation by well chosen homomorphisms over the group. However, the required number of copies of $\ket{aHG_u'}$ to find a non trivial solution to the quadratic system is now polynomial in $d$ (Theorem 5, \cite{ISS07-2}). This scaling gets worse in the characteristic of the field for larger groups, as shown for general nilpotent groups.

\subsection{An algorithm for groups of constant nilpotency class over small fields} {\label{subsection: 3.5}}
In 2023, Muhammad Imran and Gabor Ivanyos presented an adaptation of this method for nilpotent groups in \cite{Iv23}. However, this method only works for groups of constant nilpotency class $n$ and constant characteristic of the field ($p$ in the case of a $p$-group). In other words, their algorithm is exponential in $\log(|G|)$, but it is still an interesting direction to see what seems to break for bigger groups. They introduce formally the problem of computing coset states for the subgroup $HG'$ as the HSMC (Hidden Subgroup Modulo the Commutator subgroup). To solve this problem generically they describe an algorithm to hide the group $HL$, where $L$ is a subgroup of the center, by constructing a purification of the mixed state $\rho_{HL}$. In particular they derived the following lemma which gives an equivalent condition to being a purification of a hidden subgroup state:

\begin{lemma} {\label{lemma: purif}} 
    Let $H$ be a subgroup of $G$ and let $X$ be a left transversal for $H$ (i.e. a set $X \subset G$ such that $G = \displaystyle\bigsqcup_{x\in X}xH$). Then, a pure state $\ket{\psi} \in \C [G] \otimes V$, for some Hilbert space $V$, is a purification of the subgroup state $\rho_H =\frac{1}{|G|}\displaystyle\sum_{a\in G}\ket{aH} \bra{aH}$ if and only if it can be written as
    \[\ket{\psi} = \frac{1}{\sqrt{|G|}} \sum_{x \in G}\ket{x}\ket{v(x)} = \sqrt{\frac{|H|}{|G|}}\sum_{x\in X}\ket{xH}\ket{v(x)}\]
    where $v: C[G] \to V$ is a function such that the states $\ket{v(x)}$ and $\ket{v(y)}$ are equal if $x$ and $y$ are in the same left coset of $H$ and orthogonal otherwise.
\end{lemma}
In particular, one can then measure the last register of this purification to obtain a coset state $\ket{xH}$ for a uniformly random $x\in G$. This method works by generating zero sum subsequences from the states $\ket{aHL_u'} = \frac{1}{\sqrt{|L|}}\displaystyle\sum_{z \in L}\omega^{zu}\ket{aHz}$. The zero sum subsequence problem is defined like this:
\begin{definition} (Zero sum subsequence problem) \\
    Let $a_1,...,a_r \in \Z_p^n$. The zero sum subsequence problem of length $r$ over $\Z_p^n$ is the problem of finding a non empty binary string $u\in \Z_2^r$ (or equivalently a sequence) such that \[\displaystyle\sum_{i=1}^ru_i \cdot a_i =0 \mod p\]
\end{definition}
A result by Olson \cite{Ols69} implies that the minimum length $r$ for the zero sum subset problem to have a solution is $1+n(p-1)$. This is problematic as the cost of making such a state already becomes exponential in $\log(p)$, and there seem to be an unbreakable barrier to be efficient in the characteristic of the field here. There exists a deterministic algorithm that can find such a sequence with a length of size $p^{O(p\log(p))}n^{O(p \log(p))}$ (Theorem 2, \cite{Iv23}). In particular, if we consider the field to be of constant characteristic, such an algorithm is polynomial in $n$. 
Using this generation procedure, the algorithm's idea is to reduce the HSP problem to many HSMC problems. This is done by updating two subgroups $H_0$ and $K$ such that $H_0 \leq K\cap H$. The algorithm will run the HSMC procedure on the groups $(H \cap K)K'/H_0$ whenever $K/K'H_0$ is isomorphic to some $\Z_q^m, q \in \mathbb{P}$. Depending on whether $(H \cap K)K' = K$ or not, we update $H_0$ or $K$ until $H_0=K=H$ (Proposition 3, \cite{Iv23})
This algorithm is a bit complex and involves the resolution of a number of HSCM problem that is exponential in $n$. Moreover, the case when $K/K'H_0$ is not of the form $\Z_q^m$ must be treated each time. 
\\
A few natural questions arise after the progress of research:
\begin{enumerate}
    \item What stops working in the passage of nil-2 $p$-groups to general nilpotent groups?
    \item Can the Pretty Good Measurement approach be adapted for nilpotent groups as well?
    \item Can the method for general nilpotent groups be adapted to the structure of an explicit example, like $U_n(\F_p)$?
\end{enumerate}
In the following section, we explore the two first questions by studying a particular group of nilpotency class 3: $U_4(\F_p)$.

\section[Reduction of the general HSP on U4 to simpler groups]{Reduction of the general HSP on $U_4$ to simpler groups} \label{section: U4 general reduction}

An efficient algorithm for the HSP on $U_4(\F_p)$ should have a time complexity in $O(poly \log p)$. For low regimes of $p$, one can use the quantum algorithm in \cite{Iv23} for nilpotent groups of small nilpotency class over fields of small characteristic. In particular, we will suppose that $p\geq 5$, such that every element of $U_4$ is of order $p$ (this comes from the fact that $U_4$ is of nilpotency class $3 < p$).

We consider the following group with standard matrix multiplication as a group law:
\[U_{4} = \left\{
\begin{pmatrix}
1 & b_{1} & b_2 & a \\
0 & 1 & z & c_1 \\
0 & 0 & 1 & c_2 \\
0 & 0 & 0 & 1
\end{pmatrix}, (a,z,b_1,b_2,c_1,c_2)\in \Z_p^6 \right\}
\]
Semi direct product formalism allows to describe this group in two different ways:
\[U_{4} = \h_{5} \rtimes \Z_p \ ; \ U_{4} = K \rtimes \Z_p^2 \]
\[\h_{5}= \left\{
\begin{pmatrix}
1 & b_{1} & b_{2} & a \\
0 & 1 & 0 & c_{1} \\
0 & 0 & 1 & c_{2} \\
0 & 0 & 0 & 1
\end{pmatrix}, (a,b_1,b_2,c_1,c_2)\in \Z_p^5 \right\} \ ; \ K= \left\{
\begin{pmatrix}
1 & 0 & b_{2} & a \\
0 & 1 & z & c_{1} \\
0 & 0 & 1 & 0 \\
0 & 0 & 0 & 1
\end{pmatrix}, (a,z,b_2,c_1)\in \Z_p^4 \right\}\]

We note elements of $U_4$ by the 6-tuple $g=(a,z,b_1,b_2,c_1,c_2)$, 
and $[g,g']:=g\cdot g' \cdot g^{-1} \cdot g'^{-1}$ denotes the group commutator. $ K \simeq \Z_p^4, D \simeq \Z_p^2, Y\simeq \Z_p $ is abelian and $\h_5$ is the Heisenberg group of order $p^5$. $U_4$ admits the following central series:
\[U_4 \geq U_4' = \left\{
\begin{pmatrix}
1 & 0 & b_{2} & a \\
0 & 1 & 0 & c_{1} \\
0 & 0 & 1 & 0 \\
0 & 0 & 0 & 1
\end{pmatrix}, (a,b_2,c_1)\in \Z_p^4 \right\} \geq C = \left\{
\begin{pmatrix}
1 & 0 & 0 & a \\
0 & 1 & 0 & 0 \\
0 & 0 & 1 & 0 \\
0 & 0 & 0 & 1
\end{pmatrix}, a\in \Z_p^4 \right\} \geq \{I_4\}\]
where $C$ is the center of the group. \\

This section aims at proving the following result:
\begin{proposition} {\label{proposition: U_4 red}} 
    To solve the HSP on $U_4(\F_p)$, with $p$ a prime larger than 5, it is enough to solve the HSP with the promise that the hidden subgroup is of the form
    \begin{enumerate}
        \item $H=\braket{h^{(0)}}$ or $H=\{0\}$, for some unknown $h^{(0)}\in U_4$ with every diagonal element different than $0$
        \item $H=\braket{h^{(0)},h^{(1)}}$ and $H=\{h^{(1)}\}$, where the element $h^{(1)}\in U_4'$ is known and $h^{(0)}\in U_4$ has every diagonal element different than $0$
    \end{enumerate}
\end{proposition}

We give ourselves a function $f:U_{4} \to \Z_2^q$, for some integer $q$ that hides a subgroup $H \leq U_{4}$. A first approach with a small group like $U_4$ is to try to adapt Lemma \ref{lemma: hsp in A x Zp} that worked for $\h_3 \simeq \Z_p \rtimes \Z_p^2$. However, $U_4$ cannot be made of the form $A \rtimes \Z_p$ with $A$ an abelian group. To extend this method, we will study the two mentioned structures of $U_4 = K \rtimes D = \h_5 \rtimes Y$ and try to achieve the same quotient idea when possible. 

The algorithm starts by running the abelian HSP on $K$ to infer $H_1:= H \cap K$ and on the Heisenberg group $\h_5$ (one can use the implementation in \cite{KR08}) to infer $H_2 = H \cap \h_5$. The HSP algorithm on those two groups gives two generating sets:
\begin{itemize}
    \item $S_{H_1}$ a list of size $0 \leq r \leq 4$ of elements of $H_1$ such that $H_1 = \langle S_{H_1} \rangle$. We can eliminate redundant element in the list by computing the Smith normal form of the matrix whose columns are the generating elements given by the algorithm. This will give us the group invariants of $H_1$, i.e. a minimal description of $H_1$. This classical post-processing is efficient in $poly \log(p)$. Moreover, the order of $H_1$ is directly given by the length of $S_{H_1}$ after this procedure, and can be checked in constant time.
    \item $S_{H_2}$ a list of size $poly \log p$ of generating elements of $H_2$ such that $H_2 = \langle S_{H_2} \rangle$. Supposing that this list is of minimal size is possible but not necessary for our problem.
\end{itemize}

We already have the following result:
\begin{proposition} {\label{proposition: H1 not in H2 trivial}}
    The fact that $H_1 \not\subset H_2$ can be tested efficiently, and if it is the case, we can directly infer the group $H$.
\end{proposition}
\begin{proof}
    We can test if $H_1 \not\subset H_2$ in constant time as we only need to check if any element in $S_{H_1}$ has a non-zero $z$-coordinate. Then note $h \in H_1$ such an element. By Proposition \ref{prop:order-product} applied to the groups $H$ and $H_2$, $|H|\leq p|H_2|$. Now, because $h \in H \setminus H_2$, this means that $H= \langle H_2,h \rangle$. So $S_{H_2} \cup h$ is a generating set for the hidden subgroup $H$.
\end{proof}

In the following, we suppose that we cannot use this proposition, meaning that every generating element of $H_1$ has the $z$-coordinate equal to 0. In particular, it means that $H_1 \leq U_4'$ is a subgroup of the first commutator group, and is of order $1, p, p^2$ or $p^3$. Because $H_1$ is not necessarily normal in $U_4$, we must treat each case separately in hope of achieving the reduction.

\subsection[Case when H1 is of order p3 or p2]{Case when $H_1$ is of order $p^3$ or $p^2$} \label{subsection: H1 order p3 or p2}
In this section, we denote by $\begin{pmatrix} b_2 & a \\ z & c_1 \end{pmatrix} := (a,z,0,b_2,c_1,0)$ the elements of $K$. 
\\
If $H_1$ is of order $p^3$, then it is equal to $U_4'$. The quotient $U_4/U_4'$ is abelian, so we can output generators of $H$ in this case by solving an abelian HSP on the resulting quotient.
\\
If $H_1$ is of order $p^2$, then it belongs to one of the following groups:
\begin{lemma} {\label{lemma: subgroups p2 in K}}
    The subgroups of order $p^2$ of $U_4'$ are:
    \[
    \mathcal{G}_1 := \left\{\begin{pmatrix} \star & \star  \\ 0 & 0  \\ \end{pmatrix}, \begin{pmatrix} 0 & \star  \\ 0 & \star  \\ \end{pmatrix},  \left\langle \begin{pmatrix} 0 & 1  \\ 0 & 0  \\ \end{pmatrix}, \begin{pmatrix} 1 & 0  \\ 0 & u  \\ \end{pmatrix} \right\rangle, u\in \Z_p\setminus\{0\} \right\}
    \]
    \[
    \mathcal{G}_2 := \left\{\begin{pmatrix} \star & 0 \\ 0 & \star  \\ \end{pmatrix}, \left\langle \begin{pmatrix} 1 & 0 \\ 0 & 0  \\ \end{pmatrix}, \begin{pmatrix} 0 & 1  \\ 0 & u  \\ \end{pmatrix} \right\rangle, \left\langle\begin{pmatrix} 0 & 0 \\ 0 & 1  \\ \end{pmatrix}, \begin{pmatrix} u & 1  \\ 0 & 0  \\ \end{pmatrix} \right\rangle, \left\langle \begin{pmatrix} u & 1 \\ 0 & 0  \\ \end{pmatrix}, \begin{pmatrix} 0 & 1  \\ 0 & v  \\ \end{pmatrix} \right\rangle, u,v\in \Z_p \setminus \{0\} \right\}
    \]
    Moreover, $\mathcal{G}_1$ contains the normal subgroups in $U_4$ and $\mathcal{G}_2$ contains the non normal subgroups.
\end{lemma}
\begin{proof}
    Note that $U_4' \simeq \Z_p^3$, so its number of subgroups is equal to $\binom{3}{2}_p = p^2+p+1$, where $\binom{n}{k}_p=\displaystyle\prod_{i=0}^{k-1}\frac{p^n-p^i}{p^k-p^i}$ is the Gaussian binomial coefficient (see the book \cite{Led44}). The above subgroups are indeed in a number of $p^2+p+1$ and are different, so they describe all the subgroups of order $p^2$ of $U_4'$. It is then easy to check that a normal subgroup of $U_4'$ must contain the center $C$.
\end{proof}
\begin{proposition} {\label{proposition: H1 in GG1}}
    If $H_1 \in \mathcal{G}_1$, there exists an efficient quantum algorithm that describes $H$.
\end{proposition}
\begin{proof}
We show that in this case, $U_4/H_1$ is isomorphic to $\Z_p \times \h_3$. This group is itself isomorphic to $\Z_p \times (\Z_p^2 \rtimes \Z_p) \simeq \Z_p^3 \rtimes \Z_p$, with the space $\Z_p$ on the left acting trivially in the definition of the semi direct product. Then, by the result of \cite{BCD05}, one can solve the HSP on groups of the form $\Z_p^r \rtimes \Z_p$, for any $r \in \mathbb{N}$ with a quantum algorithm running in time $poly \log(p)$.
We provide isomorphisms for every group $H_1$ of $\mathcal{G}_1$ between $U_4/H_1$ and $\Z_p \times \h_3$:
\[
\begin{array}{ccccc}
    \varphi & : & U_4/\begin{pmatrix} \star & \star  \\ 0 & 0  \\ \end{pmatrix} & \to & \Z_p \times \h_3 \\
    & & (0,z,b_1,0,c_1,c_2) \cdot H_1 & \mapsto & (b_1) \times (z,c_1,c_2) \\
\end{array}
\]
\[\begin{array}{ccccc}
    \varphi & : & U_4/\begin{pmatrix} 0 & \star  \\ 0 & \star  \\ \end{pmatrix} & \to & \Z_p \times \h_3 \\
    & & (0,z,b_1,b_2,0,c_2) \cdot H_1 & \mapsto & (c_2) \times (z,b_2,b_1) \\
\end{array}
\]
\\
\[\begin{array}{ccccc}
    \varphi & : & U_4/ \left\langle \begin{pmatrix} 0 & 1  \\ 0 & 0  \\ \end{pmatrix}, \begin{pmatrix} 1 & 0  \\ 0 & v  \\ \end{pmatrix} \right\rangle & \to & \Z_p \times \h_3 \\
    & & (0,z,b_1,b_2,c_1,c_2) \cdot H_1 & \mapsto & (b_1) \times (z,c_1-vb_2,c_2) \\
\end{array}
\]
\end{proof}
Now, we consider the case when $H_1$ is not a normal subgroup of $K$, and we demonstrate the following fact:
\begin{proposition} {\label{proposition: H in GG_2}}
    If $H_1 \in \mathcal{G}_2$, then $H=H_1$
\end{proposition}
\begin{proof}
For the sake of contradiction, let us suppose that there exists some $h \in H \setminus H_1$. Write it as $h= (a,z,b_1,b_2,c_1,c_2),$ with $(b_1,c_2) \neq (0,0)$. We will cancel out the terms $b_2$ and $c_1$ by exploiting the structure of $H_1$, and then show that the condition $(b_1,c_2) \neq (0,0)$ will imply that $H$ contains the center $C$ of our group.
\\
For instance, let us suppose that $H_1= \left\langle \begin{pmatrix} 1 & u \\ 0 & 0  \\ \end{pmatrix}, \begin{pmatrix} 0 & v  \\ 0 & 1  \\ \end{pmatrix} \right\rangle, u,v\neq 0$. Note $\hat{u}, \hat{v}$ these two elements. Define $\hat{h}:= h \cdot \hat{u}^{-b_2} \cdot \hat{v}^{-c_1} = (a',z,b_1,0,0,c_2)$, where $a'= a-b_1c_1-ub_2-vc_1$. For any $k \in \Z,$ define
\[u_k := \hat{u}^k \cdot \hat{h} \cdot \hat{u}^{-k}=(a' +kc_2,z,b_1,0,0,c_2) \in H\]
\[v_k := \hat{v}^{-k} \cdot \hat{h} \cdot \hat{v}^k=(a' +kb_1,z,b_1,0,0,c_2) \in H\]
By hypothesis, $(b_1,c_2) \neq (0,0)$. If for instance, $b_1 \neq 0$, choose $k$ such that $kb_1 \equiv 1 \mod p$. Then, 
\[v_k \cdot \hat{h}^{-1} = (a' +kb_1 -a', z-z, b_1-b_1,0,0,c_2-c_2) = (1,0,0,0,0,0) \in H\]
which is a generator of $C$. Otherwise, $c_2\neq0$, so choose $k$ such that $kc_2 \equiv 1 \mod p$ such that \[u_k \cdot \hat{h}^{-1} = (a' +kc_2 -a', z-z, b_1-b_1,0,0,c_2-c_2) = (1,0,0,0,0,0) \in H\]
In the other cases of $H_1 \in \mathcal{G}_2$, the idea is the same, we simply won't have the non linear terms in $a'$, which changes nothing in the proof. 
\end{proof}

\subsection[Case when H1 is of order p]{Case when $H_1$ is of order $p$} {\label{subsection: H1 order p}}
It is easy to check that the only subgroup of $U_4'$ of order $p$ normal in $U_4$ is equal to the center $C$ of the group.
The fact of whether or not $H_1=C$ is a normal subgroup will have different effect on the coset states. We can already infer the subgroup directly in the case when $H_1 \neq H_2$:
\begin{proposition} {\label{proposition: H2 different de H1 + order p can be solved}}
    Suppose that $H_2 \neq H_1$, with $H_1$ of order $p$. Then $H=H_2$.
\end{proposition}
\begin{proof}
    We suppose that $H_2 \neq H_1$, with $H_1$ of order $p$. We can test if $H_2 \neq H_1$ by checking if every element of $S_{H_2}$ has non trivial $b_1$ or $c_2$ coordinates. We note $g=(a,0,b_1,b_2,c_1,c_2)$ an element in $H_2 \setminus H_1$ obtained this way, such that $(b_1^{(0)},c_2^{(0)}) \neq (0,0)$. $H_1$ is a non trivial cyclic group because it is of order $p$, and is a subgroup of $U_4'$. We note $H_1= \langle (a^{(1)},0,0,b_2^{(1)},c_1^{(1)},0) \rangle=\langle h^{(1)} \rangle$.
    For the sake of contradiction, suppose that $H \neq H_2$, and let $h^{(0)} \in H \setminus H_2$. It has a non-zero element as the $z$-coordinate, so by taking it to some power, we can suppose that its $z$-coordinate is 1, and we still note by $h^{(0)}$ this element:
    \[h^{(0)}=(a^{(0)},1,b_1^{(0)},b_2^{(0)},c_1^{(0)},c_2^{(0)}), \text{ with }(b_1^{(0)},c_2^{(0)}) \neq (0,0)\]
    Now, consider $[h^{(0)},g] = (\hat{a},0,0,-b_1,c_2,0) \in H_1$, where $\hat{a}\in \Z_p$. $H_1$ is cyclic, so there exists some $k \in \Z$ such that 
    $\begin{cases}
        -b_1=kb_2^{(1)} \\
        c_2=kc_1^{(1)}
    \end{cases}$. 
    In the particular case when $H_1=C$, $b_2^{(1)}=c_1^{(1)} = 0$, so that $b_1=c_2=0$, which is a contradiction with the fact that $g \in H_2 \setminus H_1$. Otherwise, $(b_2^{(1)},c_1^{(1)}) \neq (0,0)$ and the subgroup $H_1 = \braket{h^{(1)}}$ is not normal in $U_4$. Then, $k \neq 0$ as $(b_1,c_2 \neq (0,0)$ and $(b_2^{(1)},c_1^{(1)}) \neq (0,0)$
    Moreover, compute $[h^{(1)},g] = (-b_1c_1^{(1)}+b_2^{(1)}c_2,0,0,0,0,0) \in H_1 \cap C =\{0\}$
    This means that $b_1c_1^{(1)}=b_2^{(1)}c_2$. Combining this with the previous equation, we get:
    \[-kb_2^{(1)}c_1^{(1)} = kb_2^{(1)}c_1^{(1)} \iff 2kc_1^{(1)}b_2^{(1)}=0\] with $k\neq 0, c_1^{(1)}b_2^{(1)} \neq 0$ and $p \neq 2$ (we supposed that $p \geq 5$). This contradiction means that $H=H_2$. 
\end{proof}
Otherwise, the following proposition follows from Proposition \ref{prop:order-product} and the fact that an order $p^2$ group is always abelian:
\begin{proposition} {\label{proposition: form of ab groups order p2}}
    Suppose that $H_2 = H_1=\braket{h^{(1)}}$ is of order $p$. Then $H$ is either equal to $\braket{h^{(1)}}$, or is an abelian subgroup of order $p^2$ generated by $h^{(1)}$ and an element in $U_4 \setminus \h_5$.
\end{proposition}
We now treat the remaining case when $H_1$ is trivial.

\subsection[Case when H1 is trivial]{Case when $H_1$ is trivial} {\label{subsubsection: 4.2.4}}

The assumption $p\geq 5$ ensures that every non-identity element has order $p$, which yields the following proposition:
\begin{proposition} {\label{proposition: H1 trivial = H cyclic}}
    Suppose that $H_1$ is trivial. Then $H$ is a cyclic subgroup of order $p$ or the trivial group.
\end{proposition}
\begin{proof}
    For the sake of contradiction, let us suppose that $H$ is not cyclic nor trivial and that $H_1$ is trivial. By Proposition \ref{prop:order-product}, $|H| \leq p |H_2|$ and $|H| \leq p^2 |H_1|=p^2$, so $H$ must be of order $p^2$ and therefore is abelian. It is generated by two different elements $h$ and $h'$ that are not in $U_4'$. The constraint $|H| \leq p |H_2|$ imposes that $H_2$ is of order $p$ and that one of these two elements is in $H_2$. We will suppose w.l.o.g. that it is the case of $h'$. We note the two generators:
    $h =(a,z,b_1,b_2,c_1,c_2)$ and $h' =(a',z',b_1',b_2',c_1',c_2')$, with $z \neq 0,(b_1,c_2) \neq (0,0), z'=0, (b_1',c_2')\neq (0,0)$
    and those elements must commute to ensure that $H$ is abelian (it is of order $p^2$ in a $p$-group, so it is abelian). Their group commutator should be the identity element, which gives the relations:
    $\begin{cases}
        b_1'z=b_1z' =0\\ c_2'z=c_2z' =0
    \end{cases}$
    implying that $z=0$ or $(b_1',c_2')=(0,0)$. This contradiction means that $H$ is cyclic.
\end{proof}

\subsection{Details on the global algorithm}
Here, we summarize the previous steps to prove Proposition \ref{proposition: U_4 red} in the form of an algorithm:

\begin{itemize}
    \item Run the HSP algorithm on $H_1=H\cap K$ and $H_2=H \cap \h_5$ and obtain lists $S_{H_1}, S_{H_2}$. If necessary, use Smith normal form to find the group invariants of $H_1$, so that we can suppose that $S_{H_1}$ is of minimal size.
    \item Test if each element of $S_{H_1}$ is in $\h_5$ by checking their $z$-coordinate. Otherwise, return $S_{H_1} \cup S_{H_2}$ that generate $H$ by proposition \ref{proposition: H1 not in H2 trivial}.
    \item Test the order of the group $H_1$ by checking the length of $S_{H_1}$
        \begin{itemize}
        \item If $H_1$ is of order $p^3$, run the abelian HSP on $U_4/H_1 = U_4/U_4'$
        \item If $H_1$ is of order $p^2$:
            \begin{itemize}
                \item If $H_1$ is normal, solve the HSP in $U_4/H_1 \simeq \Z_p^3 \rtimes \Z_p$ with the implementation in \cite{BCD05}
                \item Otherwise, return $S_{H_1}$ that generates $H$ by Proposition \ref{proposition: H in GG_2}
            \end{itemize}
        \item If $H_1$ is of order $p$:
            \item Test if $H_1 \neq H_2$. If this is the case, then $H=H_2$
            \item Otherwise, $H$ is equal to $H_1$ (so cyclic of order $p$) or abelian of order $p^2$.
        \item If $H_1$ is trivial, $H$ is trivial or cyclic of order $p$ generated by an element in $U_4 \ K \ \h_5 $.
        \end{itemize}
\end{itemize}
This algorithm provides a way to solve the HSP on $U_4$ if the hidden subgroup $H$ is of order larger than $p^3$ and a way to test this through $H_1$ and $H_2$. Moreover, after the algorithm, we know if $H$ is cyclic or trivial or if $H$ is generated by two elements or is cyclic. This proves the reduction claimed in propositon \ref{proposition: U_4 red}.
In the next section, we explain the difficulties that rise when dealing with the two remaining cases, which are due to $U_4$ being of nilpotency class 3.

\section[HSP on U4 for order p2 and cyclic subgroups]{HSP on $U_4$ for order $p^2$ and cyclic subgroups} \label{section: U4 for p^2 and cyclic}

In this section, we describe some attempts to solve the HSP on $U_4$ for the remaining cases left by proposition \ref{proposition: U_4 red}. We show in particular the limits of the Pretty Good Measurement approach (originally derived in \cite{BCD05} for $\h_3$ and some other groups) for the nil-3 group $U_4$. This method seems ill-designed for the case when $H$ is generated by one or two elements, although we don't prove any strict limitation. This method simply fails for the case when $H$ is cyclic or trivial.

\subsection[Case when H is generated by 1 or 2 elements]{Case when $H$ is generated by 1 or 2 elements}
This particular case occurs in the algorithm presented in section \ref{subsection: H1 order p} when $H_1$ is of order $p$ and $H_1 = H_2$. Then, either $H=\braket{h^{(1)}}$, or there exist an element $h^{(0)} \in H \setminus H_2$ such that $H= \braket{h^{(1)},h^{(0)}}$. 
Note that if an algorithm can determine some non zero power of $h^{(0)}$ with high probability, then one can test if the element obtained is inside $H$. If not, then it means that $H= \braket{h^{(1)}}$ and otherwise $H=\braket{h^{(1)},h^{(0)}}$. We can thus suppose that $H$ is of the form
    \[H = \left\langle \begin{pmatrix}
1 & 0 & b_{2}^{(1)} & a^{(1)} \\
0 & 1 & 0 & c_{1}^{(1)} \\
0 & 0 & 1 & 0 \\
0 & 0 & 0 & 1
\end{pmatrix},
\begin{pmatrix}
1 & b_1^{(0)} & b_2^{(0)} & a^{(0)} \\
0 & 1 & z^{(0)} & c_1^{(0)} \\
0 & 0 & 1 & c_2^{(0)} \\
0 & 0 & 0 & 1
\end{pmatrix}
\right\rangle\]
\[ \text{Such that }  (b_1^{(0)},c_2^{(0)}) \neq (0,0), z^{(0)} \neq 0,b_1^{(0)}c_1^{(1)}=b_2^{(1)}c_2^{(0)} \]
The condition $b_1^{(0)}c_1=b_2c_2^{(0)}$ imposes that the two elements commute. From this point, one can run the quantum algorithm in \cite{BCD05} for groups of the form $\Z_p^r \rtimes \Z_p$ on the group 
\[F = \begin{pmatrix}
    1 & 0 & \star & \star \\
    0 & 1 & \star & \star \\
    0 & 0 & 1 & \star \\
    0 & 0 & 0 & 1
\end{pmatrix} \simeq K \rtimes 
\begin{pmatrix}
    1 & 0 & 0 & 0 \\
    0 & 1 & 0 & 0 \\
    0 & 0 & 1 & \star \\
    0 & 0 & 0 & 1
\end{pmatrix} \simeq \Z_p^4 \rtimes \Z_p\]
If the algorithm outputs some generator $h^{(0)} \in F \setminus U_4'$, then $H= \braket{h^{(1)},h^{(0)}}$. We suppose to be in the harder case where both coordinates $b_1^{(0)},c_2^{(0)}$ are different than 0.
\\
We now take interest in the two different cases depending on whether $h^{(1)}\in C$, resulting in the group $\braket{h^{(1)}}$ being normal in $U_4$ or not.

\subsubsection[Case when H contains the center of U4]{Case when $H$ contains the center of $U_4$} {\label{subsection: H order p2 has C}}
Because $h^{(0)}$ has every element of its diagonal different than 0, we can consider some other generator that has the $z$-coordinate equal to 1 by taking $h^{(0)}$ to some power. We can multiply this term by an appropriate element of the center such that its $a$-coordinate is 0. Then, the group $H$ is of the form 
\[H = \left\langle \begin{pmatrix}
1 & 0 & 0 & 1 \\
0 & 1 & 0 & 0 \\
0 & 0 & 1 & 0 \\
0 & 0 & 0 & 1
\end{pmatrix},
\begin{pmatrix}
1 & b_1^{(0)} & b_2^{(0)} & 0 \\
0 & 1 & 1 & c_1^{(0)} \\
0 & 0 & 1 & c_2^{(0)} \\
0 & 0 & 0 & 1
\end{pmatrix}
\right\rangle\]

In appendix \ref{annex: A}, we detail how one could adapt the Pretty good measurement approach on the Heisenberg group to this structure. This involves the resolution of a system of the form
\begin{equation}
\label{eq:sysH}
\begin{cases}
-\dfrac{\overrightarrow{\beta'}\cdot\overrightarrow{k^2}}{2}
+\overrightarrow{k}\cdot\overrightarrow{\beta}
= B_1+\dfrac{B_2}{2}, \\[8pt]
\overrightarrow{k}\cdot\overrightarrow{\beta'}=B_2, \\[8pt]
\overrightarrow{k}\cdot\overrightarrow{\gamma}=C_1, \\[8pt]
\dfrac{\overrightarrow{\gamma}\cdot\overrightarrow{k^2}}{2}
+\overrightarrow{k}\cdot\overrightarrow{\gamma'}
=C_2+\dfrac{C_1}{2}.
\end{cases}
\end{equation}
of unknown $\overrightarrow{k} \in \Z_p^4$, $(\overrightarrow{\beta},\overrightarrow{\beta'},\overrightarrow{\gamma},\overrightarrow{\gamma'})$ are random components in $\Z_p^4$ and $B_1,B_2,C_1,C_2 \in \Z_p$ are parameters of the problem. The procedure exploits a unitary realizing the transformation
$$
U\ket{\mathbf{S}_{\overrightarrow{\beta},\overrightarrow{\beta'},\overrightarrow{\gamma},\overrightarrow{\gamma'}}^{B_1,B_2,C_1,C_2}}
=
\ket{B_1,B_2,C_1,C_2},
$$

where

$$
\ket{\mathbf{S}_{\overrightarrow{\beta},\overrightarrow{\beta'},\overrightarrow{\gamma},\overrightarrow{\gamma'}}^{B_1,B_2,C_1,C_2}}
=
\frac{1}{\sqrt{\left|\mathbf{S}_{\overrightarrow{\beta},\overrightarrow{\beta'},\overrightarrow{\gamma},\overrightarrow{\gamma'}}^{B_1,B_2,C_1,C_2}\right|}}
\sum_{\overrightarrow{k}\in
\mathbf{S}_{\overrightarrow{\beta},\overrightarrow{\beta'},\overrightarrow{\gamma},\overrightarrow{\gamma'}}^{B_1,B_2,C_1,C_2}}
\ket{\overrightarrow{k}}
$$

is the uniform superposition over all the solutions of the previous system for fixed values of $B_1,B_2,C_1,C_2$. While such a transformation is well defined on the states corresponding to non-empty solution sets and can be extended to a unitary on the remaining space, its efficient implementation requires a coherent procedure for finding all the solutions of the system and preparing their uniform superposition for arbitrary values of the parameters.

The two linear equations can first be used to express two coordinates of $\overrightarrow{k}$ in terms of the other two, provided that an appropriate $2\times2$ minor of the corresponding linear system is non-zero, which occurs with high probability for uniformly random coefficients. Substituting these expressions into the two remaining equations yields two quadratic equations in two variables. Their common solutions can be found by eliminating one variable using a resultant, leading generically to a univariate polynomial of degree at most four over $\mathbb{Z}_p$. Thus, for fixed values of $B_1,B_2,C_1,C_2$, the problem reduces essentially to finding the roots of a quartic polynomial over $\mathbb{Z}_p$ and reconstructing the other coordinates. The difficulty is that the number of roots in $\mathbb{Z}_p$ depends on the parameters, and the root-finding and reconstruction procedures must be performed coherently when these parameters are in superposition. Moreover, if such a unitary can be implemented, the subsequent PGM procedure recovers the desired parameters with a success probability guaranteed to be at least $1/4$, instead of the $1/2$ guarantee obtained in the Heisenberg case. Although this does not rule out an efficient implementation, it makes the construction considerably more involved than for the Heisenberg group and we did not obtain a sufficiently simple explicit implementation of $U$. All these complications transpose as well to the other case when $H$ is not normal in $U_4$. 

\subsubsection[Case when H doesn't contain the center of U4]{Case when $H$ doesn't contain the center of $U_4$} {\label{subsection: H order p2 dont have C}}
In such a case, $(b_2^{(1)},c_1^{(1)})\neq (0,0)$. By using the commutation relation and taking the elements $h^{(1)}$ and $h^{(0)}$ to some power, we can suppose that we are working with the simpler generators
\[h_1 = \begin{pmatrix}
    1 & 0 & 1 & a^{(1)} \\
    0 & 1 & 0 & x \\
    0 & 0 & 1 & 0 \\
    0 & 0 & 0 & 1
\end{pmatrix}, h_0 := \begin{pmatrix}
    1 & 1 & 0 & a^{(0)} \\
    0 & 1 & z^{(0)} & c^{(0)}_1 \\
    0 & 0 & 1 & x \\
    0 & 0 & 0 & 1
\end{pmatrix}\]
The commuting relation imposes that the two elements share a component $x \in \Z_p$. A left transversal of this group is made of elements of the form $g=(a,0,0,c_1,c_2,z) \in U_4$. The cubic term in the first register makes the resolution even harder than the previous case. Indeed, a coset state is now of the form 
\[\begin{aligned}
    \ket{g\cdot H} &= \frac{1}{p} \displaystyle \sum_{k_0,k_1 \in \Z_p} \ket{a+k_1a^{(1)}+k_0a^{(0)}_1+k_0k_1x + {k_0 \choose 3}xz^{(0)} + {k_0 \choose 2}(b^{(0)}_2x+c^{(0)}_1)} \ket{k_0} \\
    & \qquad \ket{k_1+k_0b^{(0)}_2+{k_0 \choose 2}z^{(0)}} \ket{c_1+k_1x+k_0c^{(0)}_1+k_0xz + {k_0 \choose 2}z^{(0)}x} \ket{c_2+k_0x} \ket{z+k_0z^{(0)}}
\end{aligned}\]
Using a unitary $\mathcal{V}$ made of linear operations with the second and the third register, we can get rid of the terms in $k_0k_1x$ in the first register and in $k_0xz$ in the fourth one, but are still left with a cubic term.
\[\begin{aligned}
    \mathcal{V} \ket{g\cdot H} &= \frac{1}{p} \displaystyle \sum_{k_0,k_1 \in \Z_p} \ket{a+k_1a^{(1)}+k_0a^{(0)}_1 + \frac{k_0-k_0^3}{3}xz^{(0)} + \frac{k_0^2-k_0}{2}c^{(0)}_1 - \frac{k_0^2+k_0}{2}b^{(0)}_2x} \ket{k_0} \\
    & \qquad \ket{k_1+k_0b^{(0)}_2+\frac{k_0^2-k_0}{2}z^{(0)}} \ket{c_1+k_1x+k_0c^{(0)}_1 - \frac{k_0^2+k_0}{2}z^{(0)}x} \ket{c_2+k_0x} \ket{z+k_0z^{(0)}} \\
\end{aligned}\]
The same procedure as in subsection \ref{subsection: H order p2 has C} yields a system with 3 unknowns with a cubic term in $\overrightarrow{k}$. This makes the resolution of this equation even harder and questions how to efficiently construct the inverting unitary. Still, there is no definitive barrier here that makes PGM fails, unlike the case of the cyclic subgroup.

\subsection[The problem of random correlations for H trivial or cyclic of order p]{The problem of random correlations for $H$ trivial or cyclic of order $p$}

Note that if we have a quantum algorithm that can infer the hidden subgroup with the promise that the subgroup is cyclic and not trivial, we could still run this algorithm with the promise that the subgroup is cyclic (and can be trivial). If the procedure outputs some element $h$, we can test if $h\in H$ by one query to the oracle, and test if $H$ is cyclic or trivial. So we can suppose that the hidden subgroup $H$ is cyclic and not trivial, and generated by some $h^{(0)} = (a^{(0)},b_1^{(0)},b_2^{(0)},c_1^{(0)},c_2^{(0)},z^{(0)}) \neq I_4$. Note that we can also run the HSP algorithm with the implementation of \cite{BCD05} and of \cite{KR08} on the normal subgroups
\[K \rtimes \begin{pmatrix}
    1 & 0 & 0 & 0 \\
    0 & 1 & 0 & 0 \\
    0 & 0 & 1 & \star \\
    0 & 0 & 0 & 1
\end{pmatrix} \simeq \Z_p^4 \rtimes \Z_p \text{ ; } K \rtimes \begin{pmatrix}
    1 & \star & 0 & 0 \\
    0 & 1 & 0 & 0 \\
    0 & 0 & 1 & 0 \\
    0 & 0 & 0 & 1
\end{pmatrix} \simeq \Z_p^4 \rtimes \Z_p \text{ ; } \h_5\] 
If the result of the HSP on any of those groups is not trivial, it is a generator of $H$. Otherwise, every element of the first superdiagonal is different than 0: $b_1^{(0)},c_2^{(0)},z^{(0)} \neq 0$. In the following we suppose to be in this difficult case. We can take that element to some power $k$ such that $z^{(0)}k = 1 \mod p$, so that the z-coordinate of $h^{(0)}$ is 1. We will still denote by $h^{(0)}$ this element, and we are left to find the 5 components of $h^{(0)} = (a^{(0)},b_1^{(0)},b_2^{(0)},c_1^{(0)},c_2^{(0)},1)$.

A left transversal of the group $H$ can be formed by the elements $g=(a,b_1,b_2,c_1,c_2,0) \in U_4$. The standard method yields the coset states, for uniformly random $g=(a,b_1,b_2,c_1,c_2,0) \in \h_5$ 
\[\begin{aligned}
\ket{g\cdot H} &= \frac{1}{\sqrt{p}}\displaystyle\sum_{k\in \Z_p} \ket{g \cdot (h^{(0)})^k} \\
 &= \frac{1}{\sqrt{p}}\displaystyle\sum_{k\in \Z_p} \ket{a +k(a^0+b_1c_1^0+b_2c_2^0) + {k\choose2}(b_1^0c_1^0+b_2^0c_2^0+b_1c_2^0) + {k\choose3}b_1^0c_2^0}
 \\ &\quad \otimes \ket{b_1+kb_1^0} \otimes \ket{b_2+k(b_1+b_2^0) + {k\choose2}b_1^0} \otimes \ket{c_1+kc_1^0+{k\choose2}c_2^0} \otimes \ket{c_2+kc_2^0} \otimes \ket{k} 
\end{aligned}\]
The core problem here is that we have linear and quadratic terms in the random components $b_1$ and $b_2$ on the first register. We can apply unitary operations with the other registers to simplify a bit this expression. Although we can suppress the quadratic term in ${k\choose2} b_1c_2^0$ in the first register, there is no way to get rid of every random dependency. Noting $X_i$ the $i$-th register, the operations $X_3\leftarrow X_3-X_2X_6; \ X_1\leftarrow X_1-X_2X_4$ yield the state
\[\begin{aligned}
\ket{\mathcal{V}gH}&=
\frac{1}{\sqrt p}\sum_{k\in\mathbb Z_p}
\Bigl|
a-b_1c_1
+k(a^0+b_2c_2^0+b_1^0c_1)
-\frac{k^2+k}{2}b_1^0c_1^0
+{k\choose2}b_2^0c_2^0
-\frac{k(k^2-1)}{3}b_1^0c_2^0
\Bigr\rangle
\\
&\qquad\otimes |b_1+kb_1^0\rangle
\otimes \Bigl|b_2+kb_2^0-\frac{k^2+k}{2}b_1^0\Bigr\rangle
\\
&\qquad\otimes \Bigl|c_1+kc_1^0+\frac{k^2-k}{2}c_2^0\Bigr\rangle
\otimes |c_2+kc_2^0\rangle
\otimes |k\rangle
\end{aligned}\]
were we noted $\mathcal{V}$ the global unitary that realizes this transformation. This is the most simplified form of coset state we obtained. 
\\
If one was to try the PGM approach that worked for $\h_3$, they would need to invert a set of solutions that depends on the random components $b_2,c_1$ of $g$. This poses a severe limitation to this method, and the task of getting rid of the random correlations between $k$ and $g$ in the first component does not seem easy. For instance, one could try to apply a unitary diving the first component by the last one when this is possible. This cancels out the linear dependency in $k$ and the random outcomes $b_2,c_1$, and reduces the degree of the cubic term in the first register. However, it couples $k^{-1}$ to the other random component $a-b_1c_1$. Another attempt is to use QFTs over the full Heisenberg group $\h_5$. However, a measuring it would yield with high probability a dimension-2 irreducible representation label, and an overall state that still depends on the random phase. 
\\
The idea left out is to find a way to disentangle the first register from the rest of the state. This is equivalent to creating a coset state for the subgroup $HC$. Then, there would be no dependency in the random components in the first register and one is left with a situation akin to subsection \ref{subsection: H order p2 has C}. If we take this idea further,  one could try to produce a coset state for the subgroup $HU_4'$. Then, the quotient group $U_4/U_4'$ is abelian, and one can solve a cyclic HSP to infer the components of $h^{(0)}$ on the first super diagonal. Although there is no obvious quantum circuit that can realize this coset state generation, there exist other methods of producing purifications of $\rho_{HU_4'}$. This idea is explored in the next section for the more general group $U_n$.

\section[A general algorithm for the HSP on Un]{A general algorithm for the HSP on $U_n$} \label{section: general algo Un}

\subsection[Some notations on Un]{Some notations on $U_n$}

We now aim at solving the HSP on the more general group $U_n(\F_p)$. To exploit the nilpotency characteristic of this group, we consider the lower central series of $U_n$. For any $k \in [\![0,n-1 ]\!]$, we note \[U_{n,k}(\F_p) := [U_n,[U_n,[...[U_n,U_n]...]\] the normal subgroup of $U_n$, with the $k$ first superdiagonals equal to 0. $U_{n,k}$ is thus the $k^{th}$ normal subgroup of the descending central series of $U_n$. With this definition, we have $U_{n,0} =U_n$ and $U_{n,n-1} = \{I_n\}$. We also note $D_k$ the $k^{th}$ upper diagonal, for $k\in [\![  1,n-1]\!]$. 

To support this structure, we represent elements $u$ of $U_n$ by the list of their superdiagonals: 
$\forall u \in U_n, u :=(u_1,...,u_{n-1})$, where $\forall k \in [\![  1,n-1]\!], u_k \in D_k \simeq \Z_p^{n-k}$ such that 
\[u = I_n + \displaystyle\sum_{k=1}^{n-1}\sum_{l=1}^{n-k} u_{k,l} \ket{l}\bra{k+l}\]
The element $u_{k,l}$, for $k \in [\![  1,n-1]\!]$ and $l \in [\![  1,n-k]\!]$ is thus the component of the matrix $u$ of index $(k+l-1,k+l)$, i.e. the $l^{th}$ component of the $k^{th}$ super diagonal. The qudit $\ket{u} \in \C[{U_n}]$ is represented by the same list: 
\[\begin{aligned}
\ket{u} &= \ket{u_1} \otimes \ket{u_2} \otimes ... \otimes \ket{u_{n-1}} \\
&= \ket{u_{1,1}} \ket{u_{1,2}} ... \ket{u_{1,n-1}} \otimes \ket{u_{2,1}} ...\otimes ... \otimes \ket{u_{n-2,1}}\ket{u_{n-2,2}} \otimes \ket{u_{n-1,1}} 
\end{aligned}\]
We will indifferently denote each component $(u_k)_l = u_{k,l} \in \Z_p$. 
We note $(u_i\star v_j)_r :=u_{i,r}v_{j,r+i} \in \Z_p^{n-i-j}, \forall r \in [\![1,n-i-j]\!]$ the convolution on $n-i$ and $n-j$ tuples. The multiplication of two elements $u,v \in U_n$ is then given by 
\[\begin{aligned}
    (uv)_k &= u_k + v_k + \sum_{i=1}^{k-1} u_i \star v_{k-i} \\
    &= u_k + v_k + \sum_{i=1}^{k-1} (u_{i,r}v_{k-i,r+i})_{1 \leq r \leq n-k}
\end{aligned}\]

Below are some useful computation rules when dealing with this form:

\begin{lemma} {\label{lemma: comp rule Un}} 
Let $u,v = (u_1,...,u_{n-1}),(v_1,...,v_{n-1}) \in U_n$, and let $k \in [\![1,n-2]\!].$ note $w =u\cdot v$ and note $x= u\cdot v^{-1}$.
\begin{enumerate}
    \item $u_1 = ... = u_k = 0 \implies \forall i \in [\![1,k ]\!], w_i = v_i$ and $w_{k+1} = u_{k+1} + v_{k+1}$
    \item $v_1 = ... = v_k =0 \implies \forall i \in [\![1,k ]\!], w_i = u_i$ and $w_{k+1} = u_{k+1} + v_{k+1}$
    \item $\forall r \in \mathbb{N}, ((0,...0,u_{k},...,u_{n-1})^r)_i = \begin{array}{cc}
        0  \ \text{ if } \ i\leq k-1\\
        r \cdot u_k \ \text{ if } i=k  
    \end{array}$
\end{enumerate}
\end{lemma}

Let $H$ be a subgroup of $U_{n}$. We suppose to have access to an oracle implementing a hiding function $f$ for $H$, noted $O_f$. After applying the oracle to a uniform superposition over $U_n$ and measuring the last register, we are left with a coset state $\ket{gH}$ for a uniformly random $g \in U_n$ (we say that we have a coset generator for the group $H$). Equivalently, we have access to the mixed state \[\rho_h = \frac{1}{|U_n|}\displaystyle\sum_{g \in U_n} \ket{gH}\bra{gH} = \frac{|H|}{|U_n|}\sum_{x\in X} \ket{xH}\bra{xH}\]
where $X$ is a left transversal of $H$ in $U_n$. 

\subsection[A quantum algorithm for cyclic subgroups of Un]{A quantum algorithm for cyclic subgroups of $U_n$} \label{subsec 3.2}

We first focus on solving the HSP on $U_n$ with the promise that the hidden subgroup is cyclic. Let us suppose in this section that $H$ is generated by a single element $h^{(0)}$. For simplicity, we can suppose for now that this element is of order $p$ (this is the case if $p \geq n$). By looking at the computation rules on $U_n$, we see that the multiplication for elements in $U_{n,k}$ is just an abelian addition on the $k^{th}$ upper diagonal. An idea is thus to retrieve the components of $h^{(0)}$ diagonal by diagonal. This can be done provided we are able to produce coset states for the group $HU_{n,k}$. We have the following lemma to describe a transversal of this group:

\begin{lemma} \label{lemma: left transversal of Unk} 
    Let $h$ be a generating element for $H$ of order $p$ and let $k \in [\![0,n]\!]$. Let $i \in [\![1,n-1]\!], j \in [\![1,n-i]\!]$ be the index of the first non zero coordinate of $h$, i.e., such that $h \in U_{n,i-1}$ and $\forall l \in [\![1,j-1]\!], (h_i)_l=0$. Then, one possible left transversal of the group $HU_{n,k}$ is:
    \begin{enumerate}
        \item The elements $g=(g_1,...,g_k,0,...,0)$, where $g_l \in \Z_p^{n-l}$ if $k\leq i$, i.e. if $H \subset U_{n,k}$
        \item The elements $g=(g_1,...,g_k,0,...,0)$, where $g_l \in \Z_p^{n-l}$ and $(g_i)_j=0$ if $k>i$, i.e. if $H \not \subset H_{n,k}$.
    \end{enumerate}
\end{lemma}

The following proposition formalizes the idea that producing cosets for $HU_{n,k}$ gives a procedure to output a group generator of $H$:

\begin{proposition} {\label{proposition: coset states HUk = H if cyclic}}
    Suppose the subgroup $H$ is cyclic of order $p$. Suppose that we have access to a number $poly (n\log p)$ of coset states for $HU_{n,k}$, for any $k\in [\![0,n-1]\!]$. Then there exists a quantum algorithm that uses $poly(n\log(p))$ such states and outputs some generator of $H$.
\end{proposition}

This proposition is detailed in Annex \ref{Annex: C}. We now present a procedure adapted to the case of $U_n$ that produces purifications of the state $\rho_{HU_{n,k}}$, which are used in the previous result.

\begin{proposition} {\label{purif of HUnk}}
    There exist a quantum procedure that generates purifications of the state $\rho_{HU_{n,n-k-1}}$ for any $k$, and that takes $poly(p^k,n)$ calls to the oracle that hides the group $H$.
\end{proposition}
\begin{proof}
    The proof is by induction over $k \in [\![0,n-1]\!]$. Of course, for $k=0$, the group $HU_{n,n-0-1} = H\cdot \{0\} = H$ can be hidden by the single coset state generation. now, let $k \in [\![0,n-2]\!]$ and suppose that producing a purification for $\rho_{HU_{n,n-k-1}}$ can be done in $poly(p^k,n)$ calls to the oracle.
    \begin{itemize}
        \item $i \xleftarrow{} 0; \  Repeat \xleftarrow{} True; \ L \xleftarrow{} []$
        \item While Repeat = True: \begin{itemize}
            \item Generate a random coset state for $HU_{n,n-k-1}$: we obtain some $\ket{g_iHU_{n,n-k-1}}$, for a random $g_i \in U_n$.
            \item Create a uniform superposition over the $n-k-2$ diagonal and multiply it to the coset state to obtain the overall state:
            $\frac{1}{\sqrt{p^{k}}}\sum_{d \in D_{n-k-2}}\ket{g_i HU_{n,n-k-1}d}\ket{d}$
            \item Perform an abelian QFT over $D_{n-k-2}$ on the last register and measure an outcome $\alpha_i \in \Z_p^{k+1}$ and the state
            \[\ket{\psi_i} = \sum_{d\in D_{k+1}}\omega^{d\cdot \alpha_i} \ket{g_i H U_{n,n-k-1}d} \] $L \xleftarrow{} L + \alpha_i; \  i \xleftarrow{} i+1$
            \item Run the zero sum subset algorithm on the list $L$. If the algorithms succeeds for a sublist $I_\alpha = [\![\alpha_{i_1},..., \alpha_{i_r}]\!], Repeat \xleftarrow{} False$
        \end{itemize}
        \item Build the tensor product state 
        \[\ket{x} \ket{\psi_{i_1}}...\ket{\psi_{i_r}}\]
        \item Multiply each tensor state to the right by $x^{-1}$ and return the relabeled state 
        \[\ket{x}\ket{v(x)} = \ket{x}\ket{\psi_1 x^{-1}}...\ket{\psi_r x^{-1}}\]
    \end{itemize}
    Now, we prove that $\ket{v(x)}, x\in U_n$ satisfy the requirements for $\frac{1}{\sqrt{U_n}}\sum_{x \in U_n} \ket{x}\ket{v(x)}$ to be a purification of $\rho_{HU_{n,n-k}}$ by lemma \ref{lemma: purif}. This means that $\ket{v(x_1)}$ is equal to $\ket{v(x_2)}$ if $x_1$ and $x_2$ belong in the same left coset of $HU_{n,n-k-2}$ and are orthogonal otherwise.
    So, suppose that $x_1,x_2 \in U_n$ are in different coset of $HU_{n-k-2}$. We want to show that $\ket{\psi \cdot x_1^{-1}}$ and $\ket{\psi \cdot  x_2^{-1}}$ are orthogonal (we omit the label $i$ in $\psi_i$ as this holds for any $i$):
    \[\begin{aligned}
        \braket{\psi \cdot x_1^{-1} | \psi \cdot x_2^{-1}}&= 
        \sum_{d_1,d_2 \in D_{k+1}}\omega^{(d_1+d_2)\cdot \alpha_i}\braket{gHU_{n,n-k-1}d_1 \cdot x_1^{-1}|gHU_{n,n-k-1}d_2 \cdot x_2^{-1}} \\
        &= \sum_{d_1,d_2 \in D_{k+1}}\omega^{(d_1+d_2)\cdot \alpha_i} \sum_{h_1,h_2 \in H, u_1,u_2 \in U_{n,n-k-1}} \braket{gh_1u_1d_1 \cdot x_1^{-1}|gh_2u_2d_2 \cdot x_2^{-1}}
    \end{aligned}\]
    now, for any $d_1,d_2 \in D_{k+1}, h_1,h_2 \in H, u_1,u_2 \in U_{n,n-k-1}$, the elements $gh_1u_1d_1$ and $gh_2u_2d_2$ are in the same left coset $gHU_{n,n-k-2}$, so multiplying them by elements of different left cosets will yield different elements. So each one of them are orthogonal to each other, and the 2 states $\ket{\psi \cdot x_1^{-1}}$ and $\ket{\psi \cdot x_2^{-1}}$  are orthogonal. 
    \\ 
    Now, suppose that $x_1,x_2$ are in the same left coset of $HU_{n,n-k-2} = HU_{n,n-k-1}D_{k+1}$, and note $x_1 = x_2\cdot h\cdot u \cdot d, h \in H, u\in U_{n,n-k-1}, d\in D_{k+1}$.
    \[\begin{aligned}
        \ket{\Psi_{x_1}} &= \sum_{d_1,...,d_r \in D_{k+1}} \omega^{d_1 \cdot \alpha_1+...+d_r \cdot \alpha_r} \ket{g_1HU_{n,n-k-1} d_1\cdot (x_2hud)^{-1}} ... \ket{g_rHU_{n,n-k-1} d_r\cdot (x_2hud)^{-1}} \\
        &= \sum_{d_1',...,d_r' \in D_{k+1}} \omega^{(d_1'+d) \cdot \alpha_1+...+(d_r'+d) \cdot \alpha_r} \ket{g_1HU_{n,n-k-1}d_1'u^{-1}h^{-1}\cdot x_2^{-1}} ... \ket{g_rHU_{n,n-k-1}d_r'u^{-1}h^{-1}\cdot x_2^{-1}} \\
        &= \sum_{d_1',...,d_r' \in D_{k+1}} \omega^{d_1' \cdot \alpha_1+...+d_r' + d(\alpha_1+...+\alpha_r)} \ket{g_1HU_{n,n-k-1} \cdot d_1'(hu)^{-1}d_1'^{-1} \cdot d_1'\cdot x_2^{-1}} ...  \\
        & \qquad \qquad \qquad \qquad ... \ket{g_rHU_{n,n-k-1} \cdot d_r'(hu)^{-1}d_r'^{-1} \cdot d_r'\cdot x_2^{-1}} \\
        &=  \sum_{d_1',...,d_r' \in D_{k+1}} \omega^{d_1' \cdot \alpha_1+...+d_r' + 0} \ket{g_1HU_{n,n-k-1}d_1'\cdot x_2^{-1}} ... \ket{g_rHU_{n,n-k-1}d_r'\cdot x_2^{-1}}\\
        &=  \ket{\Psi_{x_2}}
    \end{aligned}\]
    due to the fact that $d_i (hu)^{-1} d_i^{-1} \in HU_{n,n-k-1} $. This comes from lemma \ref{lemma: comp rule Un}, the element $d_i$ having its first $k$ diagonals equal to 0, it does not change the $k^{th}$ diagonal of $hu$ and the conjugacy by $d_i$ only acts non trivially on the $U_{n,n-k-1}$ part.
    This makes of $\ket{\Psi_x}, x\in U_n$ a quantum hiding set for $HU_{n,n-k-2}$. We can hope to solve the zero sum subsequence problem in $poly (p,n)$ by the theorem $2$ of \cite{Iv23}, so the cost of producing a coset for $HU_{n,n-k-1}$ is in $poly(p^k,n)$.
\end{proof}

We can adapt this method by taking any series $Z = G_1 \trianglelefteq G_2 \trianglelefteq ... \trianglelefteq G_r$, where $Z$ is the center of $U_n$. We can in particular define $G_i$ to be the abelian group of the right $i$ coefficient of the top raw of elements of $U_n$. Then $G_{n-1}$, that we will note $V$ is the abelian groups that contains the elements of the top raw. In particular, we have the following lemma when we apply the previous method to this group:

\begin{lemma} {\label{lemma: purif of HV}}
    Purifications of the state $\rho_{HV}$, where $V$ is the abelian subgroup composed of elements of the top raw, can be produced in $poly(p^n)$ calls to the oracle for $H$.
\end{lemma}

\subsection[A quantum algorithm for any subgroup of Un]{A quantum algorithm for any subgroup of $U_n$} \label{subsec 3.3}

In this subsection, we present an alternative algorithm than the one in \cite{Iv23} for the unitriangular group $U_n$. It also involves the resolution of the zero sum subset problem to construct coset states of the product group of $H$ with $V$, the abelian subgroup of elements of the top raw, and the subroutine for cyclic subgroups presented previously. 

We write $U_{n+1}$ as the semi direct product $U_{n+1} = V \rtimes U_n$, where $V \simeq \Z_p^n$ is the top raw of the matrices of $U_{n+1}$ and $U_n$ is the $n \times n$ bloc on the bottom right part of the matrices. The group $U_n$ acts on $V$ via:
\[\forall v \in V, \forall u \in U_n, (u \cdot v,0) = (0,u)*(v,0)*(0,u)^{-1} \in V \times \{0\} \simeq V\] where $*$ denotes the usual matrix multiplication.
note that any element $v' \in V$ can be picked in the action of $u$:
\[\begin{aligned}
    (v',u)*(v,0)*(v',u)^{-1} &= (0,u)*(0,u)^{-1}*(v',u)*(v,0)*(v',u)^{-1}*(0,u)*(0,u)^{-1}  \\
    &= (0,u) * (v',0) * (v,0) * (-v',0)*(0,u^{-1}) \\
    &= (0,u)*(v'+v-v',0)*(0,u)^{-1} = (u \cdot v,0)
\end{aligned}\] 

Now, let $H$ be a subgroup of $U_{n+1}$, hidden by an oracle $O_f$ implementing a hiding function $f$. We demonstrate the following proposition:

\begin{proposition} \label{proposition: algo on Un made with rho HV}
    Provided a quantum algorithm that can solve the HSP for cyclic subgroups of $U_{n}$ and a polynomial number of purifications of $\rho_{HV}$, there exists a quantum algorithm solving the HSP for every group of $U_n$ in a polynomial number of calls to the cyclic subroutine and to purifications of $\rho_{HV}$.
\end{proposition}

With the semi-direct product structure, we have an alternative way of representing the subgroup $H$. 
\begin{proposition}
    The subgroup $H$ is determined by the triplet $(W,K,\alpha)$ where:
    \begin{enumerate}
    \item $K$ is the subgroup of $U_n$ defined as $K= \pi(H)$, where $\pi$ is the natural projection onto $U_n$: $\begin{array}{cccccc}
        \pi & : & V \rtimes U_n & \to & U_n  \\
         & & (v,u) & \mapsto &u
    \end{array}$.
    \item $W \leq V$ is defined as $W = H \cap V$. It is a subgroup of $V$ stable by the action of $K$
    \item $\alpha: K \to V$ is a map that satisfies $\forall u,u' \in K, \ \alpha(u\cdot u') - u \cdot \alpha(u') - \alpha(u) \in W$, where for any $u\in U_n, v \in V, u\cdot v := (0,u)\cdot(v,0)\cdot (0,u^{-1})$ denotes the action of $U_n$ on $V$. If we consider $\tilde{\alpha} : K \to V/W$, then $\tilde{\alpha}$ follows the definition of a cocycle.
    \end{enumerate}
    Moreover, $H = \{(w+\alpha(k),k), w\in W, k \in K\}$, and the map $\alpha$ is unique modulo $W$, i.e. two subgroups having the same $W, K$, and cocycle $\tilde{\alpha}$ are equal.
\end{proposition}
\begin{proof}
    First, define $W$ and $K$ as stated in the proposition. It is clear that those are two subgroups of respectively $V$ and $U_n$. Let us show that $W$ is stable by $K$. For any $k \in K, w\in W$, for any $v\in V$, the element $k \cdot w  = (v,k)(w,0)(v,k)^{-1}\in V$. Moreover, $k \in \pi(H)$, so there exist an element $v_k\in V$ such that $(v_k,k) \in H$. now, as $H$ is stable by taking the inverse and by operations, $k \cdot w = (v_k,k)(w,0)(v_k,k)^{-1} \in H$. So $W$ is stable by the action of $K$. \\
    now, we motivate the existence of a cocycle. For any $k\in K= \pi(H)$, there exists some $\alpha(k) \in V$ such that $(\alpha(k),k) \in H$. $H$ is stable my multiplication, so for any $k,k' \in U_n, \ (\alpha(k),k)\cdot(\alpha(k'),k') = (\alpha(k)+k\cdot \alpha(k'),kk') \in H$. We also have the element $(\alpha(kk'),kk') \in H$. Multiplying the first one by the inverse of the second one, we get 
    \[\begin{aligned}
        (\alpha(k)+k\cdot \alpha(k'),kk')\cdot (\alpha(kk'),kk')^{-1} &= (\alpha(k) +k\cdot \alpha(k') + kk' \cdot (-(kk')^{-1} \cdot\alpha(kk')),kk'(kk')^{-1}) \\ &= (\alpha(k)+k\cdot \alpha(k') -\alpha(kk'),0) \in H \cap V = W
    \end{aligned}\]
    From here, it is clear that for any $k\in K, w\in W, \ (w+\alpha(k),k) = (w,0)\cdot (\alpha(k),k))\in H$. now, take any element $h=(v,k) \in H$. $(\alpha(k),k) \cdot h^{-1} = (\alpha(k) +k \cdot (-k^{-1}\cdot v), kk^{-1}) = (\alpha(k)-v, 0) \in H \cap V = W$. We can note $w:= \alpha(k)-v$ so that $h = (\alpha(k)-w,k)$ is of the expected form, and $H =\{(w+\alpha(k),k), w\in W, k \in K\}$. \\
    Finally, take two different $\alpha_1, \alpha_2$ of $H$. 
    \[\begin{aligned}
        (\alpha_1(k),k)\cdot (\alpha_2(k),k)^{-1} &= (\alpha_1(k) +k \cdot (-k \cdot \alpha_2(k),kk^{-1}) \\
        &= (\alpha_1(k)-\alpha_2(k),0) \in H \cap V = W \\
    \end{aligned}\]
    So there exist $w_k \in W$ such that $\alpha_1(k)= \alpha_2(k) +W$, and if we consider the quotient application $\tilde{\alpha_1} : U_n \to V/W$ and $\tilde{\alpha_2}$, we indeed have that $\tilde{\alpha_1}= \tilde{\alpha_2}$
\end{proof}
We now have the objective to find the triplet $(W,K,\tilde{\alpha})$ to determine $H$. We can obtain $W$ already by running the abelian HSP on $V$.
\begin{lemma} {\label{lemma: W}}
    We can efficiently infer the subgroup $W = H \cap V$
\end{lemma}
The space $K$ is already a bit harder to infer, but the following lemma is immediate 
\begin{lemma} {\label{lemma: K}} 
    Provided we are able to solve the HSP for any subgroup of $U_n$ given a quantum procedure and that we can generate coset states for the group $HV$, there is a quantum algorithm that uses a polynomial number of copies of $\ket{HV}$ and 1 call to the quantum procedure for $U_n$ and that outputs generators of the subgroup $K = \pi(H)$
\end{lemma}
\begin{lemma} {\label{lemma: alpha}} 
    Provided we have access to a poly log number of generators of the groups $K$ and $W$ and to a quantum algorithm for cyclic subgroups of $U_{n+1}$, there is a quantum algorithm that takes a polynomial number of calls to the cyclic subroutine and computes the cocycle $\tilde{\alpha}$ such that $H=(W,K,\tilde{\alpha})$.
\end{lemma}
\begin{proof}
    We write $K=\braket{k_1,...,k_r}$. Our goal is to find the values of a cocycle $\tilde{\alpha}$ on each of the $k_i$. So let $k_i$ be one of those generators. We can apply the standard method on the group $M_i := V/W \times \braket{k_i}$. This will hide the subgroup $H_i = H \cap M_k = \{(\tilde{\alpha}(k_i^x),k_i^x), x\in [\![0,p^{m_i}-1]\!]\}$, where $p^{m_i}$ is the order of the element $k_i$. Then, $H_i$ is a cyclic subgroup of $U_{n+1}$ of order $p^{m_i}$. By applying the subroutine for cyclic subgroups of $U_{n+1}$, we can compute the cocycle on the value $\tilde{\alpha}(k_i)$. This is enough to describe $\tilde{\alpha}$ over $K$ due to the cocycle relation:
    For any $k \in K$, note $k=k_{i_1}k_{i_2} ... k_{i_r}$, where $k_{i_j} \in \{k_1,...,k_r\}$. Then, \[\tilde{\alpha}(k) = k_{i_1} \cdot \tilde{\alpha}(k_{i_2} ... k_{i_r}) + \tilde{\alpha}(k_{i_1}) = ... = \displaystyle\sum_{j=1}^r k_{i_1} ...k_{i_{j-1}} \cdot \tilde{\alpha}(k_{i_j})\]
\end{proof}
Proposition \ref{proposition: algo on Un made with rho HV} is then immediate with an induction on Lemmas \ref{lemma: W}, \ref{lemma: K} and \ref{lemma: alpha}. This proposition and the use of the result of Section \ref{subsec 3.2} allow to derive this final result:
\begin{proposition}
    There exist a quantum algorithm that can infer the subgroup $H=(W,K,\tilde\alpha)$ in time $poly(p^n)$.
\end{proposition}

\section{Conclusion} \label{section: conclusion}

Research on the extension of the Hidden Subgroup Problem to non abelian group is quite hard, even for other cases than the dihedral and the symmetric group that catch most of researchers attention. The last genuine extension lasts from 2007 with nil-2 p-groups \cite{ISS07-2}. Upper unitriangular groups are thus very natural HSP instances to consider to pursue this direction. However, irreducible representations of general unitriangular groups is "wild" and makes the use of general Quantum Fourier Transform very limited. This poses a great challenge to find efficient algorithms that bypass the standard method.
\\
In this paper, some progress was still made on small groups like $U_4$. In particular, we derived a quantum algorithm that can compute the order of the hidden subgroup and provide a list of generators in logarithmic time in $p$ if the subgroup is of order larger than $p^3$ (Proposition \ref{proposition: U_4 red}). For remaining cases, we tried to adapt the PGM approach from the Heisenberg group resolution. This method doesn't seem very natural and is quite delicate when the group is of order $p^2$. The system of equation to solve becomes very complex quite fast, with quartic equations appearing for the simplest instance of this problem (Section \ref{subsection: H order p2 has C}). This doesn't mean that this method stops working for this group, but rather suggests harder and harder subproblems if we were to try out the same reduction for groups like $U_5$, $U_6$ and so on.
\\
However, the adaptation of the cyclic HSP algorithm for the Heisenberg group to \(U_4\) encounters a fundamental difficulty arising from the random phase terms introduced by the central coordinate of the coset states. Exploring different efficiently implementable QFTs over the group did not provide a way to eliminate or control these phases. Exploiting the Abelian behavior of the first superdiagonal therefore appears to be a particularly natural alternative. We explored this direction to derive a quantum algorithm for $U_n$ by adapting methods used in \cite{Iv23}. The structure of $U_{n+1} = V \rtimes U_n$ allowed a more direct way to retrieve group generators. Although this doesn't constitute an improvement of the original method in terms of complexity, it is a bit simpler and only involves the use of abelian QFTs.
\\
These obstacles leave several directions open. For \(U_4\), the difficulties encountered in extending the Heisenberg group techniques highlight precisely where the additional non-Abelian structure becomes relevant. This work makes the difficulty of the problem explicit for nilpotent $p$-groups of class above 3. Understanding whether the dependence on the nilpotency class can be overcome thus remains an interesting open direction. 
For bigger groups like $U_n$, the reduction to zero-sum subsequence problems represents one possible way of combining several coset states and is not necessarily the only one. An efficient HSP algorithm should thus make use of a different framework than PGM and exploit the group's particular structure. One could for instance consider a filtration $U_n = \gamma_1 \trianglerighteq \gamma_2 \trianglerighteq \dots \trianglerighteq \gamma_n$ and solve successive quotient HSP. However, such an approach has its limits for cyclic subgroups where the phenomenon of random correlations appear. Exploiting the particular structure of $U_n$ for cyclic subgroups to avoid zero sum subset problem reductions does seem to be the bottleneck of this problem.

\section*{Acknowledgements}

The author would like to thank Prof. Hadi Salmasian - University of Ottawa - for their supervision, helpful discussions, and guidance throughout this work.
The author also thanks Quasar Research Group in Ottawa for providing a stimulating research environment. The author also thanks Mitacs GRA for the funding of the project and their continuous support.

This research was carried out as part of a research internship for the Master 2 QMI at Institut Polytechnique de Paris. 

This work was supported by the Mitacs Canadian program \textit{Globalink Research Awards - Quantum Stream Initiative}.

\vspace{0.6cm}

\begin{minipage}{0.5\textwidth}
        \centering
        \includegraphics[width=\textwidth]{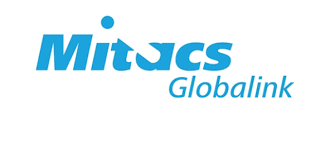}
    \end{minipage}

\newpage
\printbibliography

\newpage
\appendix
\appendixpage
\addappheadtotoc

\section[Adaptation of the PGM approach for order p2 or p subgroups of U4]{Adaptation of the PGM approach for order $p^2$ or $p$ subgroups of $U_4$}
\label{annex: A}
In this problem, the hidden subgroup $H$ is of the form obtained in Section \ref{subsection: H order p2 has C}: 
\[H = \left\langle \begin{pmatrix}
1 & 0 & 0 & 1 \\
0 & 1 & 0 & 0 \\
0 & 0 & 1 & 0 \\
0 & 0 & 0 & 1
\end{pmatrix},
\begin{pmatrix}
1 & b_1^{(0)} & b_2^{(0)} & 0 \\
0 & 1 & 1 & c_1^{(0)} \\
0 & 0 & 1 & c_2^{(0)} \\
0 & 0 & 0 & 1
\end{pmatrix}
\right\rangle\]
It is easy to check that a left transversal for $H$ is given by elements of the form $g=(0,b_1,b_2,c_1,c_2,0) \in U_4$. A random coset state generated by an element $g=(0,b_1,b_2,c_1,c_2,0)$ for $H$ is given by
\[\begin{aligned}
\ket{g\cdot H} &= \frac{1}{\sqrt{p}}\displaystyle\sum_{c\in C}\sum_{k\in \Z_p} \ket{g \cdot (h^{(0)})^k \cdot c} \\
 &= \frac{1}{\sqrt{p}}\displaystyle\sum_{c \in C}\sum_{k\in \Z_p} \ket{c +k(b_1c_1^0+b_2c_2^0) + {k\choose2}(b_1^0c_1^0+b_2^0c_2^0+b_1c_2^0) + {k\choose3}b_1^0c_2^0}
 \\ &\quad \otimes \ket{b_1+kb_1^0} \otimes \ket{b_2+k(b_1+b_2^0) + {k\choose2}b_1^0} \otimes \ket{c_1+kc_1^0+{k\choose2}c_2^0} \otimes \ket{c_2+kc_2^0} \otimes \ket{k} \\
 &= \frac{1}{\sqrt{p}}\displaystyle\sum_{c \in C}\ket{c}\sum_{k\in \Z_p} \ket{b_1+kb_1^0} \otimes \ket{b_2+k(b_1+b_2^0) + {k\choose2}b_1^0} \otimes \ket{c_1+kc_1^0+{k\choose2}c_2^0} \otimes \ket{c_2+kc_2^0} \otimes \ket{k} 
\end{aligned}\]
after a change of variable. Now, we can discard the first register that has no entanglement with the rest and obtain the state:
\[\begin{aligned}
\ket{\psi} &= \frac{1}{\sqrt{p}}\displaystyle\sum_{k\in \Z_p} \ket{b_1+kb_1^0} \otimes \ket{b_2+k(b_1+b_2^0) + {k\choose2}b_1^0} \otimes \ket{c_1+kc_1^0+{k\choose2}c_2^0} \otimes \ket{c_2+kc_2^0} \otimes \ket{k} 
\end{aligned}\]
In the following, we try to adapt the PGM approach to infer the missing components of $h^{(0)}$, but this is a more delicate situation than the Heisenberg case in Annex \ref{annex: A}. Note that we have a term in $kb_1$ in the second register that depends on the random coset. We can discard this dependency by using linear operations with the first and the last register to obtain the state:
\[\begin{aligned}
\ket{\psi} &= \frac{1}{\sqrt{p}}\displaystyle\sum_{k\in \Z_p} \ket{b_1+kb_1^0} \otimes \ket{b_2+k(b_1+b_2^0) + {k\choose2}b_1^0 - k \times (b_1 +kb_1^{(0)})} \otimes \ket{c_1+kc_1^0+{k\choose2}c_2^0} \\
& \qquad \qquad \qquad \otimes \ket{c_2+kc_2^0} \otimes \ket{k} \\
&= \frac{1}{\sqrt{p}}\displaystyle\sum_{k\in \Z_p} \ket{b_1+kb_1^0} \otimes \ket{b_2+kb_2^0 - \frac{k^2+k}{2}b_1^0 } \otimes \ket{c_1+kc_1^0+{k\choose2}c_2^0} \otimes \ket{c_2+kc_2^0} \otimes \ket{k} 
\end{aligned}\]

We can now apply an abelian QFT to the first 4 registers of the state $\ket{\psi}$:
\[
\begin{aligned}
F_{\Z_p}^{\otimes 4}\otimes Id\,|\psi\rangle
&=
\frac{1}{\sqrt{p^5}}
\sum_{k,\beta,\beta',\gamma,\gamma'\in\Z_p}
\omega^{\beta(b_1+kb_1^0)+ 
\beta'\!\left(b_2+kb_2^0 - \frac{k^2+k}{2}b_1^0 \right)
+\gamma\!\left(c_1+kc_1^0+{k\choose2}c_2^0\right)
+\gamma'(c_2+kc_2^0)
} \\
& \qquad \qquad \qquad \qquad \qquad
\ket{\beta,\beta',\gamma,\gamma',k}
\end{aligned}
\]
Now, measure the components $\beta,\beta',\gamma,\gamma'$ in the first 4 registers to obtain a state $\ket{\psi^{\beta,\beta',\gamma,\gamma'}}$. The tensor product of a number $r$ of such states after the outcomes $\overrightarrow{\beta},\overrightarrow{\beta'},\overrightarrow{\gamma},\overrightarrow{\gamma'} \in \Z_p^r$ is, up to global phase,
\[
\begin{aligned}
\ket{\Psi^{\overrightarrow{\beta},\overrightarrow{\beta'},\overrightarrow{\gamma},\overrightarrow{\gamma'}}} &=
\frac{1}{\sqrt{p^{r}}}
\sum_{\overrightarrow{k} \in\Z_p^r}
\omega^{\overrightarrow{k}\overrightarrow{\beta}b_1^0 +
\overrightarrow{\beta'}\!\left(\overrightarrow{k}b_2^0-\frac{\overrightarrow{k^2}+\overrightarrow{k}}{2}b_1^0\right)
+\overrightarrow{\gamma}\!\left(\overrightarrow{k}c_1^0+\frac{\overrightarrow{k^2}-\overrightarrow{k}}{2}c_2^0\right)
+\overrightarrow{k}\overrightarrow{\gamma'}c_2^0
} \  \ket{\overrightarrow{k}} \\
&= \frac{1}{\sqrt{p^{r}}}
\sum_{\overrightarrow{k} \in\Z_p^r}
\omega^{b_1^0[\overrightarrow{k}\overrightarrow{\beta} -\frac{\overrightarrow{k^2}+\overrightarrow{k}}{2}\overrightarrow{\beta^2}] + b_2^0 \overrightarrow{k} \overrightarrow{\beta'} + c_2^0[\overrightarrow{k} \overrightarrow{\gamma'} +\frac{\overrightarrow{k^2}-\overrightarrow{k}}{2}\overrightarrow{\gamma}] +c_1^0 \overrightarrow{k} \overrightarrow{\gamma}
} \  \ket{\overrightarrow{k}} \\
&=
\sum_{B_1,B_2,C_1,C_2 \in \Z_p}
\omega^{b_1^0B_1 + b_2^0B_2 + c_1^0C_1 + c_2^0C_2} \ket{\mathbf{S}_{\overrightarrow{\beta},\overrightarrow{\beta'},\overrightarrow{\gamma},\overrightarrow{\gamma'}}^{B_1,B_2,C_1,C_2}}
\end{aligned}
\]
where $\mathbf{S}_{\overrightarrow{\beta},\overrightarrow{\beta'},\overrightarrow{\gamma},\overrightarrow{\gamma'}}^{B_1,B_2,C_1,C_2}$ denotes the set of elements $\overrightarrow{k} \in \Z_p^r$ such that 
\begin{subequations} {\label{eq: sysH has p -1}}
    \begin{align}
    -\overrightarrow{\beta'} \frac{\overrightarrow{k^2}+\overrightarrow{k}}{2} + \overrightarrow{k}\overrightarrow{\beta}&= B_1 \\ 
     \overrightarrow{k}\overrightarrow{\beta'} &= B_2\\
     \overrightarrow{k}\overrightarrow{\gamma} &= C_1\\
     \overrightarrow{\gamma} \frac{\overrightarrow{k^2}-\overrightarrow{k}}{2} + \overrightarrow{k}\overrightarrow{\gamma'} &= C_2
     \end{align}
\end{subequations}
For any set $X$, recall that the state $\ket{X} = \frac{1}{\sqrt{|X|}}\sum_{x\in X} \ket{x}$ denotes a uniform superposition over the said set, so $\ket{\mathbf{S}_{\overrightarrow{\beta},\overrightarrow{\beta'},\overrightarrow{\gamma},\overrightarrow{\gamma'}}^{B_1,B_2,C_1,C_2}}$ is a uniform superpositions of the solutions of the equation. As there are 4 equations, one should pick $r=4$ to have a constant number of solutions. For fixed measurement outcomes
$(\overrightarrow{\beta},\overrightarrow{\beta'},
\overrightarrow{\gamma},\overrightarrow{\gamma'})$, define the map

$$
\Phi:\mathbb{Z}_p^4\longrightarrow\mathbb{Z}_p^4,
\qquad
\overrightarrow{k}\longmapsto(B_1,B_2,C_1,C_2),
$$

where $(B_1,B_2,C_1,C_2)$ are the four values obtained from the left-hand sides of \eqref{eq: sysH has p -2}. Then

$$
\mathbf{S}^{B_1,B_2,C_1,C_2}
=
\Phi^{-1}(B_1,B_2,C_1,C_2).
$$

Consequently, the non-empty sets $\mathbf{S}^{B_1,B_2,C_1,C_2}$ are precisely the fibers of $\Phi$ and form a partition of $\mathbb{Z}_p^4$:

$$
\bigsqcup_{\substack{B_1,B_2,C_1,C_2\in\mathbb{Z}_p\\
\mathbf{S}^{B_1,B_2,C_1,C_2}\neq\varnothing}}
\mathbf{S}^{B_1,B_2,C_1,C_2}
=
\mathbb{Z}_p^4.
$$

This immediately implies that the states associated with two distinct non-empty fibers are orthogonal:

$$
\left\langle
\mathbf{S}^{B_1,B_2,C_1,C_2}
\middle|
\mathbf{S}^{B'_1,B'_2,C'_1,C'_2}
\right\rangle
=0
$$

whenever
$(B_1,B_2,C_1,C_2)\neq(B'_1,B'_2,C'_1,C'_2)$.

The PGM approach motivates the construction of a unitary $U$, depending on the measurement outcomes
$(\overrightarrow{\beta},\overrightarrow{\beta'},
\overrightarrow{\gamma},\overrightarrow{\gamma'})$, such that

$$
U
\ket{\mathbf{S}_{\overrightarrow{\beta},\overrightarrow{\beta'},
\overrightarrow{\gamma},\overrightarrow{\gamma'}}^{B_1,B_2,C_1,C_2}}
=
\ket{B_1,B_2,C_1,C_2}
$$

for every non-empty fiber, and defined everywhere else in a way that this is a unitary. Hence, the issue is not the existence of such a unitary, but whether it can be implemented efficiently.

A natural way of obtaining such an implementation would be to construct an efficient coherent procedure which, given $(B_1,B_2,C_1,C_2)$, determines all the solutions of \eqref{eq: sysH has p -2} and prepares their uniform superposition. The inverse of such a state-preparation procedure, combined with the reversible evaluation of $\Phi$, would then implement the desired transformation.

This explains our choice of $r=4$. The variable
$\overrightarrow{k}$ then belongs to $\mathbb{Z}_p^4$, while the system provides four constraints parametrized by $(B_1,B_2,C_1,C_2)$. After using the two linear constraints to eliminate two coordinates, the problem reduces generically to the intersection of two plane conics. Provided that these conics have no common component, Bézout's theorem bounds their number of intersection points over the algebraic closure by four, and therefore

$$
\left|
\mathbf{S}_{\overrightarrow{\beta},\overrightarrow{\beta'},
\overrightarrow{\gamma},\overrightarrow{\gamma'}}^{B_1,B_2,C_1,C_2}
\right|
\leq 4.
$$

Thus, for generic measurement outcomes, each non-empty fiber contains at most four solutions. This constant bound is particularly useful for the coherent preparation of the corresponding uniform superposition. Notice, however, that the bound does not follow merely from choosing $r=4$: exceptional choices of the measured coefficients may lead to degenerate systems in which the two conics share a component.

We now investigate whether these solutions can be computed efficiently. Using the two linear equations, the system can first be rewritten as
\begin{subequations}{\label{eq: sysH has p -2}}
    \begin{align}
    -\overrightarrow{\beta'} \frac{\overrightarrow{k^2}}{2}
    + \overrightarrow{k}\overrightarrow{\beta}
    &= B_1 +\frac{B_2}{2},
    {\label{eq: 2A}}\\
    \overrightarrow{k}\overrightarrow{\beta'}
    &= B_2,
    {\label{eq: 2B}}\\
    \overrightarrow{k}\overrightarrow{\gamma}
    &= C_1,
    {\label{eq: 2C}}\\
    \overrightarrow{\gamma} \frac{\overrightarrow{k^2}}{2}
    + \overrightarrow{k}\overrightarrow{\gamma'}
    &= C_2 + \frac{C_1}{2}.
    {\label{eq: 2D}}
    \end{align}
\end{subequations}

Equations \eqref{eq: 2B} and \eqref{eq: 2C} can be used to express $k_1$ and $k_2$ as affine functions of $k_3$ and $k_4$. Assuming

$$
\delta=\beta'_1\gamma_2-\beta'_2\gamma_1\neq0,
$$

we obtain

$$
k_1
=
\delta^{-1}
\left[
k_3(\gamma_3\beta'_2-\gamma_2\beta'_3)
+k_4(\gamma_4\beta'_2-\gamma_2\beta'_4)
+(B_2\gamma_2-C_1\beta'_2)
\right],
$$

and

$$
k_2
=
\delta^{-1}
\left[
k_3(\gamma_1\beta'_3-\gamma_3\beta'_1)
+k_4(\gamma_1\beta'_4-\gamma_4\beta'_1)
+(C_1\beta'_1-B_2\gamma_1)
\right].
$$

Since $\beta'$ and $\gamma$ are uniformly distributed, $\delta\neq0$ with high probability; more precisely,

$$
\Pr(\delta\neq0)
=
\left(1-\frac{1}{p}\right)
\left(1-\frac{1}{p^2}\right).
$$

Substituting these expressions into \eqref{eq: 2A} and \eqref{eq: 2D} yields two quadratic equations in $(k_3,k_4)$. In particular, mixed terms in $k_3k_4$ generally appear, so the system is no longer reducible to two independent univariate quadratic equations. It can instead be viewed as the intersection of two conics,

$$
f_{B_1,B_2,C_1,C_2}(k_3,k_4)=0,
\qquad
g_{B_1,B_2,C_1,C_2}(k_3,k_4)=0.
$$

Eliminating $k_3$ (or $k_4$) by taking the resultant of these two quadratic polynomials gives, in the generic case, a univariate polynomial of degree at most four. Hence the original system can in principle be solved by finding the roots of a polynomial of bounded degree over $\mathbb{Z}_p$ and reconstructing the remaining coordinates. This also makes the bound of four solutions transparent in the non-degenerate case.

For fixed values of $B_1,B_2,C_1,C_2$, this procedure is straightforward to implement computationally. The difficulty relevant to the quantum algorithm is stronger: the transformation must be performed coherently for all values of $(B_1,B_2,C_1,C_2)$ occurring in superposition. One must therefore determine the number of solutions, compute them according to a consistent convention, and coherently prepare

$$
\frac{1}{\sqrt{
|\mathbf{S}^{B_1,B_2,C_1,C_2}|
}}
\sum_{\overrightarrow{k}\in
\mathbf{S}^{B_1,B_2,C_1,C_2}}
\ket{\overrightarrow{k}},
$$

while also treating the parameter values for which the fiber is empty or the polynomial system becomes degenerate.

In principle, elimination techniques such as resultants or Gröbner bases provide algebraic descriptions of these solutions. However, the resulting symbolic expressions in
$(B_1,B_2,C_1,C_2)$ rapidly become very heavy, and an explicit expression alone does not establish an efficient coherent implementation. In particular, one must account for finite-field root finding, the parameter-dependent number of roots, exceptional degeneracies, and the reversible preparation of a uniform superposition over the resulting solutions.

We therefore do not obtain an impossibility result from this analysis. Rather, it shows that the direct extension of the solution used for the Heisenberg group $\mathbb{H}_3$ loses much of its simplicity already in this case. Although the polynomial system has bounded degree and appears classically tractable for fixed parameters, we did not find a sufficiently simple and uniform reversible procedure to justify an efficient implementation of the corresponding PGM unitary.

Moreover, when attempting to extend this strategy to larger unitriangular groups, one expects systems involving more variables and higher-degree elimination polynomials. The complexity of the symbolic description and of coherent root handling may therefore increase with the dimension and nilpotency class. This observation is a limitation of the present approach rather than a lower bound: it does not rule out the existence of a more efficient construction based on a different parametrization or algebraic method.

If such a unitary $U$ can be implemented efficiently, retrieving the components $(b_1^0,b_2^0,c_1^0,c_2^0)$ becomes straightforward using the PGM method. For fixed measurement outcomes $(\overrightarrow{\beta},\overrightarrow{\beta'},\overrightarrow{\gamma},\overrightarrow{\gamma'})$, the sets
$\mathbf{S}_{\overrightarrow{\beta},\overrightarrow{\beta'},\overrightarrow{\gamma},\overrightarrow{\gamma'}}^{B_1,B_2,C_1,C_2}$
form a partition of $\mathbb{Z}_p^4$. Therefore, regrouping the terms of the state according to these fibers gives
$$
\ket{\Psi^{\overrightarrow{\beta},\overrightarrow{\beta'},\overrightarrow{\gamma},\overrightarrow{\gamma'}}}
=
\frac{1}{p^2}
\sum_{B_1,B_2,C_1,C_2\in\mathbb{Z}_p}
\sqrt{
\left|
\mathbf{S}_{\overrightarrow{\beta},\overrightarrow{\beta'},\overrightarrow{\gamma},\overrightarrow{\gamma'}}^{B_1,B_2,C_1,C_2}
\right|
}
\omega_p^{b_1^0B_1+b_2^0B_2+c_1^0C_1+c_2^0C_2}
\ket{
\mathbf{S}_{\overrightarrow{\beta},\overrightarrow{\beta'},\overrightarrow{\gamma},\overrightarrow{\gamma'}}^{B_1,B_2,C_1,C_2}
},
$$

where terms corresponding to empty fibers are understood to have zero amplitude.

Applying the unitary $U$ such that

$$
U
\ket{
\mathbf{S}_{\overrightarrow{\beta},\overrightarrow{\beta'},\overrightarrow{\gamma},\overrightarrow{\gamma'}}^{B_1,B_2,C_1,C_2}
}
=
\ket{B_1,B_2,C_1,C_2}
$$

for every non-empty fiber yields

$$
U
\ket{\Psi^{\overrightarrow{\beta},\overrightarrow{\beta'},\overrightarrow{\gamma},\overrightarrow{\gamma'}}}
=
\frac{1}{p^2}
\sum_{B_1,B_2,C_1,C_2\in\mathbb{Z}_p}
\sqrt{
\left|
\mathbf{S}_{\overrightarrow{\beta},\overrightarrow{\beta'},\overrightarrow{\gamma},\overrightarrow{\gamma'}}^{B_1,B_2,C_1,C_2}
\right|
}
\omega_p^{b_1^0B_1+b_2^0B_2+c_1^0C_1+c_2^0C_2}
\ket{B_1,B_2,C_1,C_2}.
$$

Applying an inverse abelian Fourier transform over each of the four registers then gives

$$
\begin{aligned}
\ket{\Psi_{\overrightarrow{\beta},\overrightarrow{\beta'},\overrightarrow{\gamma},\overrightarrow{\gamma'}}'}
&=
{(F_{\mathbb{Z}_p}^{-1})}^{\otimes 4}
U
\ket{\Psi^{\overrightarrow{\beta},\overrightarrow{\beta'},\overrightarrow{\gamma},\overrightarrow{\gamma'}}}
\\
&=
\frac{1}{p^4}
\sum_{b_1',b_2',c_1',c_2'\in\mathbb{Z}_p}
\sum_{B_1,B_2,C_1,C_2\in\mathbb{Z}_p}
\sqrt{
\left|
\mathbf{S}_{\overrightarrow{\beta},\overrightarrow{\beta'},\overrightarrow{\gamma},\overrightarrow{\gamma'}}^{B_1,B_2,C_1,C_2}
\right|
}
\\
&\qquad\qquad
\times
\omega_p^{
B_1(b_1^0-b_1')
+B_2(b_2^0-b_2')
+C_1(c_1^0-c_1')
+C_2(c_2^0-c_2')
}
\ket{b_1',b_2',c_1',c_2'}.
\end{aligned}
$$

We now measure the four registers. The amplitude corresponding to the desired outcome
$(b_1^0,b_2^0,c_1^0,c_2^0)$ is

$$
\frac{1}{p^4}
\sum_{B_1,B_2,C_1,C_2\in\mathbb{Z}_p}
\sqrt{
\left|
\mathbf{S}_{\overrightarrow{\beta},\overrightarrow{\beta'},\overrightarrow{\gamma},\overrightarrow{\gamma'}}^{B_1,B_2,C_1,C_2}
\right|
},
$$

since all the phases cancel for this outcome. Hence, conditionally on the measured values
$(\overrightarrow{\beta},\overrightarrow{\beta'},\overrightarrow{\gamma},\overrightarrow{\gamma'})$, the probability of recovering the correct tuple is

$$
\mathbb{P}_{\mathrm{success}}
=
\frac{1}{p^8}
\left(
\sum_{B_1,B_2,C_1,C_2\in\mathbb{Z}_p}
\sqrt{
\left|
\mathbf{S}_{\overrightarrow{\beta},\overrightarrow{\beta'},\overrightarrow{\gamma},\overrightarrow{\gamma'}}^{B_1,B_2,C_1,C_2}
\right|
}
\right)^2.
$$

In the non-degenerate case considered above, every fiber contains at most four elements:

$$
0
\leq
\left|
\mathbf{S}_{\overrightarrow{\beta},\overrightarrow{\beta'},\overrightarrow{\gamma},\overrightarrow{\gamma'}}^{B_1,B_2,C_1,C_2}
\right|
\leq 4.
$$

For every integer $s\in{0,1,2,3,4}$, we have

$$
\sqrt{s}\geq\frac{s}{2}.
$$

Moreover, since the non-empty fibers form a partition of $\mathbb{Z}_p^4$,

$$
\sum_{B_1,B_2,C_1,C_2\in\mathbb{Z}_p}
\left|
\mathbf{S}_{\overrightarrow{\beta},\overrightarrow{\beta'},\overrightarrow{\gamma},\overrightarrow{\gamma'}}^{B_1,B_2,C_1,C_2}
\right|
=
p^4.
$$

It follows that

$$
\begin{aligned}
\mathbb{P}_{\mathrm{success}}
&=
\frac{1}{p^8}
\left(
\sum_{B_1,B_2,C_1,C_2\in\mathbb{Z}_p}
\sqrt{
\left|
\mathbf{S}_{\overrightarrow{\beta},\overrightarrow{\beta'},\overrightarrow{\gamma},\overrightarrow{\gamma'}}^{B_1,B_2,C_1,C_2}
\right|
}
\right)^2
\\
&\geq
\frac{1}{p^8}
\left(
\frac{1}{2}
\sum_{B_1,B_2,C_1,C_2\in\mathbb{Z}_p}
\left|
\mathbf{S}_{\overrightarrow{\beta},\overrightarrow{\beta'},\overrightarrow{\gamma},\overrightarrow{\gamma'}}^{B_1,B_2,C_1,C_2}
\right|
\right)^2
\\
&=
\frac{1}{p^8}
\left(\frac{p^4}{2}\right)^2
\\
&=
\frac{1}{4}.
\end{aligned}
$$

Therefore, whenever the measurement outcomes are such that every fiber has size at most four, the correct value $(b_1^0,b_2^0,c_1^0,c_2^0)$ is recovered with constant probability larger than $\frac{1}{4}$. We notice a weaker result than in the Heisenberg case, where the average probability was $\frac{1}{2}$, which already highlights an increasing difficulty with this problem.

\section[Derivation of the algorithm for cyclic subgroups of Un]{Derivation of the algorithm for cyclic subgroups of $U_n$}
\label{Annex: C}

This section aims at proving the following claim:
\begin{proposition}
    Suppose the subgroup $H \leq U_n$ is cyclic of order $p$. Suppose that we have access to a number $poly (n\log p)$ of coset states for $HU_{n,k}$, for any $k\in [\![0,n-1]\!]$. Then there exists a quantum algorithm that uses $poly(n\log(p))$ such states and outputs some generator of $H$.
\end{proposition}

The idea is to infer a generating element of $H$ diagonal by diagonal, and to "uncompute" the passed diagonal at each new step using its previous known components. Observing the subgroup $HU_{n,k}$ will precisely help dealing with the $k^{th}$ diagonal.
    The proof is by finite induction over $k \in [\![1,n-1 ]\!].$ The induction hypothesis of rank $k$ is that we can efficiently recover components $h_1,...h_k$ such that $H$ is generated by some element $h=(h_1,...,h_k,h_{k+1},...,h_{n-1})$. 
    \\
    Suppose that we recovered the first $k-1$ components of some generator $h$ of $H$, with $1\leq k\leq n-1$.
    \\
    In a first step, we will suppose that each one of these components is equal to 0, meaning that $h \in U_{n,k}$. The coset state $\ket{gHU_{n,k}}$ is given by some uniformly random $g=(g_1,...,g_k,0,...,0) \in U_n$:
    \[\begin{aligned}
    \ket{g HU_{n,k}} &= \frac{1}{\sqrt{p}\sqrt{|HU_{n,k}|}}\sum_{x =0}^{n-1}\sum_{u \in U_{n,k}}\ket{g\cdot h^x \cdot u} \\
    &= \frac{1}{\sqrt{p}\sqrt{|HU_{n,k}|}}\sum_{x =0}^{n-1}\sum_{u \in U_{n,k}}\ket{(g_1,...,g_{k},0,...,0)\cdot (0,...,0, x \cdot h_k, ...) \cdot (0,...,0,u_{k+1},...,u_{n-1})} \\
    &= \frac{1}{\sqrt{p}\sqrt{|HU_{n,k}|}}\sum_{x =0}^{n-1}\sum_{u \in U_{n,k}}\ket{g_1,...,g_{k-1}} \ket{ g_k +x \cdot h_k} \ket{(g\cdot h^x \cdot u)_{k+1}, ..., (g\cdot h^x \cdot u)_{n-1}}\\
    &= \frac{1}{\sqrt{p}} \ket{g_1,...,g_{k-1}}\sum_{x =0}^{n-1}\ket{(g_k + x \cdot h_k} \frac{1}{\sqrt{|U_{n,k|}}}\sum_{v \in U_{n,k}}\ket{v_{k+1}, ..., v_{n-1}}
    \end{aligned}\]
    Where we did the change of variable "$v = g\cdot h^x \cdot u$", which is indeed a bijection when $x$ is fixed as $u$ ranges over $U_{n,k}$. The first $k-1$ and the last $n-k-1$ registers now hold no information on the subgroup and we can discard them to obtain the state:
    \[\ket{\psi_g} = \frac{1}{\sqrt{p}}\sum_{x =0}^{n-1}\ket{g_k + x \cdot h_k}\]
    note that this is the same state we start from when we aim at solving the abelian HSP on $\Z_p^{n-k}$. For this reason, the same method will work and allow us to retrieve $h_k$ with high probability in an efficient number of queries to $O_{f_k}$. We can use the abelian implementation in \cite{Chi25} for instance.
    \\
    Now, we treat the case when there exists some component $(h_i)_j \neq 0$. By lemma \ref{lemma: left transversal of Unk}, the coset states of $HU_{n,k}$ are given by the representatives $g=(g_1,...,g_k,0,...,0) \in U_n$ such that the $j^{th}$ coordinate of $g_i$ is equal to 0.  The coset state $\ket{gHU_{n,k}}$ is given by some uniformly random $g=(g_1,...,g_k,0,...,0) \in U_n$: 
    \[\begin{aligned}
    \ket{g HU_{n,k}} &= \frac{1}{\sqrt{p}\sqrt{|HU_{n,k}|}}\sum_{x =0}^{n-1}\sum_{u \in U_{n,k}}\ket{g\cdot h^x \cdot u} \\
    &= \frac{1}{\sqrt{p}\sqrt{|HU_{n,k}|}}\sum_{x =0}^{n-1}\sum_{u \in U_{n,k}}\ket{(g_1,...,g_{k},0,...,0)\cdot (0,...,0, x \cdot h_i, ...) \cdot (0,...,0,u_{k+1},...,u_{n-1})} \\
    &= \frac{1}{\sqrt{p}\sqrt{|HU_{n,k}|}}\sum_{x =0}^{n-1}\sum_{u \in U_{n,k}}\ket{g_1,...,g_{i-1}} \ket{ g_i +x \cdot h_i} \ket{(g\cdot h^x)_{i+1} ,...,(g\cdot h^x)_{k})} \\
    & \qquad \qquad \qquad  \qquad \otimes \ket{(g\cdot h^x \cdot u)_{k+1}, ..., (g\cdot h^x \cdot u)_{n-1}}\\
    &= \frac{1}{\sqrt{p}} \ket{g_1,...,g_{i-1}}\sum_{x =0}^{n-1}\ket{ g_i +x \cdot h_i} \ket{(g\cdot h^x)_{i+1} ,...,(g\cdot h^x)_{k})} \frac{1}{\sqrt{|U_{n,k|}}}\sum_{v \in U_{n,k}}\ket{v_{k+1}, ..., v_{n-1}}
    \end{aligned}\]
    note that the first $i-1$ and the last $n-k-1$ register hold no information on the group. This time, we will keep them under the respective labels $\ket{g'}$ and $\ket{\phi}$ to make the next operations more natural.
    \[\begin{aligned}
    \ket{\psi_g} &= \frac{1}{\sqrt{p}} \sum_{x =0}^{n-1}\ \ket{g'}\ket{ g_i +x \cdot h_i} \ket{(g\cdot h^x)_{i+1} ,...,(g\cdot h^x)_{k})}\ket{\phi} \\
    &= \frac{1}{\sqrt{p}} \sum_{x =0}^{n-1} \ket{g'}\ket{(g_i)_1,...,(g_i)_{j-1}, x,(g_i)_{j+1}+x\cdot (h_i)_{j+1},...,(g_i)_{n-i}+x\cdot (h_i)_{n-i}} \\
    &\qquad \qquad \otimes \ket{(g\cdot h^x)_{i+1} ,...,(g\cdot h^x)_{k})} \ket{\phi}
    \end{aligned}\]
    At this point, we will uncompute the diagonals $i, i+1,...,k-1$ using the known values of $h$, using a series of operations consisting of multiple multiplications with the register $\ket{x}$. We define the following unitaries:
    \begin{enumerate}
        \item A multiplication operator $M \in \mathcal{U}(\C^{U_n} \otimes \C^{U_n})$ such that $\forall u_1,u_2 \in U_n, M\ket{u_1,u_2} = \ket{u_1 \cdot u_2, u_2}$
        \item A power operator $P \in \mathcal{U}(\C^p\otimes \C^{U_n})$ such that $\forall k\in \Z_p, u\in U_n, P\ket{k,u} = \ket{k,u^k}$
        \item A Swap operator $SWAP \in \mathcal{U}(\C^p \otimes \C^{U_n})$, for any $l < m \in [\![1,n-1]\!]$ such that $\forall \ket{x} \in \C^p, \forall u=(u_1,...,u_{n-1}) \in U_n, SWAP \ket{x}\ket{u} = \ket{u} \ket{x}$
        \item An inversion operator $INV \in \mathcal{U}(\C^{U_n})$ such that $\forall u \in U_n, INV\ket{u} = \ket{u^{-1}}$
    \end{enumerate}
    Before applying those, we will copy the $j^{th}$ register in an ancilla using a controlled-X gate ($X$ being the generalization on $\C^p$ of the qubit case: $\forall \ket{x} \in \C^p, X \ket{x} = \ket{x \oplus 1}$), and we will prepare a copy of ${\eta}:=(h_1,...,h_{k-1},0,...,0)=(0,...,0,h_i,...,h_{k-1},0,...,0)$ in an auxiliary ancilla as well. \\
    Although the procedure seems complicated, the core idea is to just multiply the inverse of $\ket{\eta}$ (which contains the data we already gathered about the generator) to the state. The operations make sure to cancel correctly each power of $\ket{h}$ and to get back to a state uncorrelated from those powers in the end.
    \[\begin{aligned}
        \ket{\Psi_g} &= CX_j^{|U_n|+1}(\ket{\psi_g}\ket{0} \ket{\eta}) \\
        &= \frac{1}{\sqrt{p}} \sum_{x =0}^{n-1} \ket{g'}\ket{(g_i)_1,...,(g_i)_{j-1}, x,(g_i)_{j+1}+x\cdot (h_i)_{j+1},...,(g_i)_{n-i}+x\cdot (h_i)_{n-i}} \\
    &\qquad \qquad \otimes \ket{(g\cdot h^x)_{i+1} ,...,(g\cdot h^x)_{k})} \ket{\phi} \otimes \ket{x} \otimes \ket{\eta}
    \end{aligned}\]

    \[\begin{aligned}
    \ket{\Psi_g} &\xrightarrow{\;I_{U_n} \otimes INV \otimes I_{U_n}\;} &\frac{1}{\sqrt{p}} \sum_{x =0}^{n-1} \ket{g'}\ket{(g_i)_1,...,(g_i)_{j-1}, x,(g_i)_{j+1}+x\cdot (h_i)_{j+1},...,(g_i)_{n-i}+x\cdot (h_i)_{n-i}} \\
    && \otimes \ket{(g\cdot h^x)_{i+1} ,...,(g\cdot h^x)_{k})} \ket{\phi} \otimes \ket{-x} \otimes \ket{\eta} \\
    &\xrightarrow{\;I_{U_n} \otimes P\;} &\frac{1}{\sqrt{p}} \sum_{x =0}^{n-1} \ket{g'}\ket{(g_i)_1,...,(g_i)_{j-1}, x,(g_i)_{j+1}+x\cdot (h_i)_{j+1},...,(g_i)_{n-i}+x\cdot (h_i)_{n-i}} \\
    && \otimes \ket{(g\cdot h^x)_{i+1} ,...,(g\cdot h^x)_{k})} \ket{\phi} \otimes \ket{-x} \otimes \ket{\eta^{-x}} \\
    &\xrightarrow{\;I_{U_n} \otimes SWAP\;} & \frac{1}{\sqrt{p}} \sum_{x =0}^{n-1} \ket{g'}\ket{(g_i)_1,...,(g_i)_{j-1}, x,(g_i)_{j+1}+x\cdot (h_i)_{j+1},...,(g_i)_{n-i}+x\cdot (h_i)_{n-i}} \\
    && \otimes \ket{(g\cdot h^x)_{i+1} ,...,(g\cdot h^x)_{k})} \ket{\phi} \otimes \ket{\eta^{-x}} \otimes \ket{-x}\\
    &\xrightarrow{\;M \otimes I_p\;} & \frac{1}{\sqrt{p}} \sum_{x =0}^{n-1} \ket{g'}\ket{(g_i)_1,...,(g_i)_{j-1}, x - x,(g_i)_{j+1}+x\cdot (h_i)_{j+1} -x\cdot (\eta_i)_{j+1},...} \\
    && \otimes \ket{...,(g_i)_{n-i}+x\cdot (h_i)_{n-i} - x\cdot (\eta_i)_{n-i}} \otimes \ket{(g\cdot h^x\cdot \eta^{-x})_{i+1} ,...,(g\cdot h^x \cdot \eta^{-x})_{k})} \\ && \otimes \ket{\phi} \otimes \ket{\eta^{-x}} \otimes \ket{-x}\\
    &\xrightarrow{\;I_{U_n} \otimes I_{U_n} \otimes INV\;}& \frac{1}{\sqrt{p}} \sum_{x =0}^{n-1} \ket{g'}\ket{(g_i)_1,...,(g_i)_{j-1}, 0,(g_i)_{j+1},...,(g_i)_{n-i}} \otimes \ket{(g)_{i+1} ,...,(g)_{k-1})} \\ 
    && \otimes \ket{(g\cdot h^x\cdot \eta^{-x})_k } \ket{\phi} \otimes \ket{\eta^{-x}} \otimes \ket{x}\\
    &\xrightarrow{\;I_{U_n} \otimes SWAP^{-1}\;}& \frac{1}{\sqrt{p}} \sum_{x =0}^{n-1} \ket{g'}\ket{g_i,g_{i+1} ,...,g_{k-1})}\ket{(g\cdot (h \eta^{-1})^x)_k } \ket{\phi} \otimes \ket{x} \otimes \ket{\eta^{-x}}\\
    &\xrightarrow{\;I_{U_n} \otimes P\;}& \frac{1}{\sqrt{p}} \sum_{x =0}^{n-1}\ket{g_1,...,g_{k-1})}\ket{(g_k)_1 + x \cdot (h_k)_1, ..., (g_k)_{n-k} + x \cdot (h_k)_{n-k}} \ket{\phi} \otimes \ket{x} \otimes \ket{\eta}
    \end{aligned}\]

    We can now discard the registers that do not depend on $x$ and that hold no information on the group. We obtain the final state:
    \[\ket{\Phi_g} = \frac{1}{\sqrt{p}} \sum_{x =0}^{n-1}\ket{(g_k)_1 + x \cdot (h_k)_1, ..., (g_k)_{n-k} + x \cdot (h_k)_{n-k}} \ket{x}\]
    The abelian method for the HSP allows to efficiently recover the component $h_k$.
    \\
    Each step in the induction uses basic multiplication between elements and solves an abelian HSP on $\Z_p^{n-k}$. The overall cost of the algorithm is thus in $poly(n\log(p))$ calls to the coset state generation for the group $HU_{n,k}$.

\end{document}